\documentclass[12pt]{article}

\usepackage[margin=1in]{geometry}

\usepackage{amssymb}
\usepackage{amsmath}
\usepackage{bm}

\usepackage{lineno}

\usepackage{xcolor}
\usepackage{hyperref}
\hypersetup{
    colorlinks=true,
    linkcolor={blue!50!black},
    citecolor={blue!50!black},
    urlcolor={blue!80!black}
}

\usepackage{float}
\usepackage{tabularray}
\usepackage[normalem]{ulem}
\usepackage{graphicx}
\UseTblrLibrary{booktabs}
\UseTblrLibrary{siunitx}
\NewTableCommand{\tinytableDefineColor}[3]{\definecolor{#1}{#2}{#3}}

\usepackage{booktabs}
\usepackage{caption}
\usepackage{subcaption}
\usepackage{longtable}
\usepackage{colortbl}
\usepackage{array}
\usepackage{multirow}

\usepackage{anyfontsize}
\usepackage[T1]{fontenc}
\usepackage{lmodern}

\usepackage{titlesec}

\usepackage{natbib}
\usepackage[resetlabels]{multibib}
\newcites{appendix}{Supplementary References}

\usepackage{tikz}
\usepackage{rotating}

\usepackage{authblk}

\title{A Quasi-Experiment comparing the health of unhoused people who
have and have not experienced an eviction in King County, WA}

\author[1]{Ihsan Kahveci}
\author[2]{Timothy A. Thomas}
\author[1]{Nathalie E. Williams}
\author[3]{Janelle Rothfolk}
\author[3]{Cathea Carey}
\author[4]{Paul Hebert}
\author[4]{Amy Hagopian}
\author[1,*]{Zack W. Almquist}

\affil[1]{Department of Sociology, University of Washington, Seattle, WA}
\affil[2]{Institute of Governmental Studies, University of California, Berkeley, CA}
\affil[3]{King County Regional Homelessness Authority, Seattle, WA}
\affil[4]{School of Public Health, University of Washington, Seattle, WA}
\affil[*]{Corresponding author: \href{mailto:zalmquist@uw.edu}{zalmquist@uw.edu}}

\date{}

\begin{document}

\maketitle

\begin{abstract}
\noindent
Home eviction poses a significant threat to housing stability, a critical determinant of health. This study examines the relationship between eviction and health and substance use within the unhoused population of King County, Washington. Using a sample of 1,106 individuals experiencing homelessness, we employed a quasi-experimental design to compare the health outcomes of those who have experienced eviction with those who have not. Our findings reveal eviction is associated with an 8.3\% point increase (SE = 0.039) in the likelihood of reporting poor general health and an 9.5\% increase (SE = 0.032) in substance use disorder. No significant effect was found for mental health outcomes. While these results highlight the severe health risks linked to eviction, further research with more precise estimates is necessary to better understand long-term effects. These findings contribute to the growing evidence of how home eviction undermines the well-being of vulnerable populations.
\end{abstract}


\clearpage

\nolinenumbers

\section{Introduction}

\noindent
Lack of access to stable housing is one of the most important drivers of homelessness in the United States \citep{colburn_homelessness_2022}. When a person is experiencing housing precarity, one of the worst forms of housing loss is eviction. Qualitative work has shown that the short timeline leaves little room for recovery, frequently resulting in severed support networks, family separations, and cascading economic hardships such as job loss, court fees, and moving costs \citep{desmond_evictions_2015}. Quantitative studies have further demonstrated that people who experience eviction typically have worse outcomes than those who do not \citep{vasquez-vera_threat_2017}. The trauma of eviction has been linked to negative health consequences in both cross-sectional and longitudinal studies \citep{hoke_health_2021}.  

In this study, we examined whether eviction as a pathway to homelessness is associated with worse health outcomes than those who have not experienced eviction but are also experiencing homelessness. 
Importantly, people experience homelessness in various ways: some are unsheltered (e.g., living outside or in vehicles), others stay in emergency shelters, and some reside in temporary arrangements such as couch-surfing or staying with friends or family. Throughout this paper, we refer to all of these groups collectively as ``unhoused'' or ``people experiencing homelessness.''

The relationship between eviction and homelessness is often cyclical, as eviction leads to long-term instability in housing \citep{rutan_concentrated_2021}. This occurs because individuals who have been evicted face significant barriers to securing stable housing, including damaged credit, a lack of rental history, and discriminatory practices by landlords \citep{desmond_eviction_2012, garcia_many_2021}. Formal eviction records can persist on credit histories for up to seven years, further compounding these challenges. However, these individual struggles are deeply intertwined with systemic factors. Rising rents, stagnant wages, and inadequate social safety nets drive widespread housing instability, disproportionately affecting low-income households \citep{colburn_homelessness_2022}. 

Eviction is particularly concentrated in marginalized communities, which face overlapping disadvantages such as racial discrimination, housing segregation, and limited access to resources \citep{vasquez-vera_threat_2017}. These structural inequities underscore how housing instability is both a cause and consequence of systemic barriers, perpetuating cycles of poverty and displacement. Understanding eviction as an individual event and a systemic phenomenon is crucial to addressing its root causes.

Despite progress in documenting eviction’s impacts, key gaps remain. Formal eviction records, such as those tracked by Princeton’s Eviction Lab, estimate an annual average of 3.6 million cases between 2008 and 2018 \citep{gromis_estimating_2022}. However, these data exclude informal evictions, which are estimated to occur 5.5 times as often as formal evictions, accounting for approximately 85\% of all eviction cases \citep{gromis_estimating_2021}.
 Informal evictions involving intimidation, neglect of repairs, or illegal tactics like changing locks often escape official datasets. This study uniquely addresses this gap by capturing both formal and informal evictions, offering a broader perspective on eviction experiences of the unhoused population. Traditional datasets like household surveys often exclude unhoused individuals, limiting the scope of their findings \citep{tourangeau_hard--survey_2014}. Our sample includes respondents across diverse housing situations—from unsheltered and sheltered to temporarily and permanently housed—enabling estimation of eviction’s direct role on health within the broader spectrum of housing instability.

To understand the relationship between eviction and health, this study focuses on King County, WA, the 12th largest county in the United States, with around 2.2 million people, and the fourth largest population of people experiencing homelessness, with over 16,868 in 2024 \citep{kcrha_pit_2025}. In 2023, \cite{almquist_innovating_2024} surveyed  1,106 individuals experiencing homelessness in King County, WA. Data collection included HUD-mandated demographic questions, social network measures, and detailed health and needs assessments. Using a quasi-experimental design, we compare individuals experiencing homelessness due to eviction with those who are homeless for other reasons. This design enables us to assess the impact of eviction on physical health and substance use while controlling for housing status. Our findings highlight that 24\% (95\% CI: 19\%--30\%) of the unhoused population in King County reported a history of eviction. For context, eviction filings in the county represented approximately 1.8\% of renting households over a 12-month period (September 2023 to August 2024) \citep{thomas_washington_2024}, suggesting that individuals who experience homelessness bear a vastly disproportionate burden of eviction.

The remainder of the paper is organized as follows. In Section~\ref{sec:background}, we provide background on eviction and its health consequences. Section~\ref{sec:methods} describes our empirical strategy, including data sources and analytic methods. Section~\ref{sec:results} presents the main findings. In Section~\ref{sec:discussion}, we situate these results in the broader literature, consider their policy implications, and discuss key limitations, including the use of cross-sectional data and potential sources of measurement error.

\section{Background}\label{sec:background}
This paper frames eviction as a discrete event that precedes homelessness and examines its specific health impacts on this population by comparing individuals who have experienced eviction to those who have not. Although the health consequences of eviction may overlap with those associated with homelessness or broader housing insecurity, its suddenness and severity likely intensify these effects and exacerbate existing vulnerabilities. In King County, WA, for instance, a tenant has 14 days after a notice to pay back rent or vacate the premises \citep{washington_state_legislature_residental_nodate}. Including court proceedings and, if warranted, Sheriff enforcement, the time to evict a tenant typically concludes within a month. Although hard to measure, anecdotal evidence suggests informal evictions follow a similarly rapid timeline \citep{fowle_evading_2024}.

\subsection{Health and housing insecurity}

Housing insecurity, encompassing both eviction and homelessness, profoundly affects general health, mental health, and substance use—dimensions that align with the three survey outcomes analyzed in this study. Research links these experiences to a range of adverse health outcomes, including poorer self-rated health, heightened exposure to infections, and increased rates of chronic conditions \citep{niccolai_eviction_2019}. Cardiovascular risks and mortality are significantly elevated among individuals experiencing housing instability \citep{arcaya_effects_2013, graetz_impacts_2024, martin_adults_2019}. Mental health problems, such as higher rates of depression, anxiety, and suicide, are driven by the acute stress and instability associated with losing housing \citep{burgard_housing_2012, cagney_onset_2014}. Substance use patterns often emerge as coping mechanisms under these stressful conditions, with studies noting increased smoking, alcohol consumption, and drug use among those facing housing insecurity \citep{mulia_economic_2014, murphy_housing_2013}.

While many studies examine housing insecurity in broad terms, eviction stands out for its abruptness, social stigma, and financial disruption—stressors that may trigger acute health risks \citep{desmond_eviction_2012, grainger_its_2021, hoke_health_2021}. Homelessness likely mediates part of this relationship, as the transition from housing loss to unsheltered living introduces new risks and amplifies existing vulnerabilities \citep{kushel_toward_2023, garcia_homelessness_2024}. Yet eviction itself can generate acute pressures—such as involuntary displacement and the disruption of social networks—that contribute directly to health deterioration. Rather than subsuming eviction within a broader continuum of housing instability, this study examines it as a distinct event that amplifies health vulnerabilities. By comparing evicted and non-evicted individuals using a quasi-experimental design within a sample of people experiencing homelessness, we isolate health risks associated with eviction from those attributable to homelessness alone.

\section{Data and methods}\label{sec:methods}

This study uses a representative sample of individuals aged 18 and older experiencing homelessness in King County, WA, in 2023 (see \citealp{almquist_innovating_2024}). The survey was part of a large-scale collaboration between the King County Regional Homelessness Authority and the University of Washington to improve the HUD-mandated point-in-time (PIT) biennial count \citep{kcrha_pit_2025}. The PIT count includes two separate counts: (1) a ``sheltered'' count of individuals in HUD-recognized emergency shelters, typically a ``bed'' count, and (2) a one-night count of unsheltered people living in places not designated for human habitation (e.g., cars, encampments, or the street). In many jurisdictions, the one-night count is augmented by a post-enumeration survey. This is typically a convenience sample of people experiencing homelessness in the area, not necessarily limited to those who were unsheltered on the night of the count. This approach has been criticized for under-counting the population of people experiencing unsheltered homelessness, offering limited demographic detail, and incurring high operational costs. \citep{kushel_toward_2023, tsai_annual_2022}. In response,  \citet{almquist_innovating_2024} adapted classic respondent-driven sampling (RDS) methods \citep{heckathorn_respondent-driven_1997} to provide a general framework for estimating the size and composition of the population of people experiencing unsheltered homelessness. These measures are now used by the King County Continuum of Care (the local jurisdiction for care of the unhoused community in King County, WA) and provide the core basis for the data used in \citet{almquist_innovating_2024}.

The dataset comprises 1,106 respondents recruited using a RDS, a social network-based sampling strategy that relies on peer referrals. Eligibility criteria included adult age, providing informed consent, and identifying as experiencing homelessness, according to the U.S. Department of Housing and Urban Development (HUD) definition of homelessness\footnote{HUD defines homelessness as an individual or family who lacks a fixed, regular, and adequate nighttime residence, meaning: (i) has a primary nighttime residence that is a public or private place not meant for human habitation; (ii) is living in a publicly or privately operated shelter designated to provide temporary living arrangements (including congregate shelters, transitional housing, and hotels and motels paid for by charitable organizations or by federal, state, and local government programs); or (iii) is exiting an institution where they have resided for 90 days or less and who resided in an emergency shelter or place not meant for human habitation immediately before entering that institution \citep{us_department_of_housing_and_urban_development_homeless_2011}.}. The study also included individuals with past experiences of homelessness, even if their current housing status ranged from unsheltered to stably housed (see Table~\ref{tbl:housing}). Interviews were conducted in English and Spanish, with real-time translation services available for participants requiring additional language support. For the purposes of this analysis, we excluded 246 respondents who were randomly assigned to a separate survey experiment testing for social desirability bias and were not asked about eviction history. This yielded a final analytical sample of 860 respondents.

\subsection{Primary exposure: Eviction history}
The primary exposure variable in this study is the respondent's eviction history. Participants were asked whether they had ever been evicted from a rental property. Those who responded ``yes'' were classified as having experienced an eviction; follow-up questions assessed whether the eviction was formal (i.e., court-ordered) and the respondent's age at the time of first eviction. Approximately 24\% (95\% CI: 19\%--30\%)\footnote{The unadjusted sample proportion was 25\%. The survey-based estimates using the RDS-II method \citep{volz_probability_2008} are reported in the text.} of the sample reported a history of eviction, a rate notably higher than the county-wide eviction filing rate of 1.8\% of renting households over a 12-month period \citep{thomas_washington_2024}. Among those evicted, 46\% experienced a formal, court-ordered eviction. Nationally, informal evictions are estimated to occur 5.5 times as often as formal evictions \citep{gromis_estimating_2021}, meaning formal evictions account for roughly 15\% of all cases---substantially lower than the rate observed in our sample. Seattle-area data similarly suggest that informal evictions far exceed formal ones \citep{gromis_estimating_2021}. We calculated time since eviction by subtracting the respondent's reported age at first eviction from their current age. As shown in Figure~\ref{fig:evict_time_cat}, most evictions occurred well before the survey (median = 12.5 years; mean = 14.6 years). This temporal distance is consistent with viewing eviction as a prior exposure. See~\ref{variables} for additional details and full distributions of the eviction measures.
\subsection{Health outcomes: General health, mental health \& substance use}

To assess respondents' health, we focused on three key metrics: self-rated health, mental health, and substance use. The first measure, self-rated health, was derived from a question asking respondents to rate their health as ``poor,'' ``fair,'' ``good,'' or ``excellent.'' For analysis, this variable was recoded into a binary format, where one indicates ``poor'' or ``fair'' health (53\% of respondents), and zero indicates ``good'' or ``excellent'' health (47\%). Sensitivity analyses using alternative categorizations yielded similar results, confirming the robustness of self-rated health as a predictor of various health outcomes, including morbidity, physiological function, and mortality \citep{idler_self-rated_1997, jylha_what_2009}.

For mental health, respondents were asked whether they considered themselves to have a serious mental health disorder. While this measure relies on self-identification rather than clinical diagnosis, it is the standard approach required by HUD in the PIT methodology \citep{us_department_of_housing_and_urban_development_hmis_nodate}. Respondents could choose to skip the question or answer ``do not know.'' In our analysis, missing values were excluded, and responses were categorized as ``Yes'' or ``No'' (proportions available in Table~\ref{tbl:evict}). Similarly, respondents were asked whether they identified as having a substance use disorder. As with the mental health question, this measure was self-reported, with options for ``skip'' and ``do not know'' included. The resulting percentages are reported in Table~\ref{tbl:evict}, with further details available in the~\ref{variables}.

\begin{table}
\fontsize{10.0pt}{14.4pt}\selectfont
\caption{Baseline characteristics of evicted and non-evicted subjects  in analytical sample.}
\label{tbl:evict}
\centering
\begin{tabular*}{\linewidth}{@{\extracolsep{\fill}}lccc}
\toprule
\textbf{Characteristic} & \textbf{Non-Evicted}\textsuperscript{\textit{1}} & \textbf{Evicted}\textsuperscript{\textit{1}} & \textbf{SMD}\textsuperscript{\textit{2}} \\ 
& (N = 629) & (N = 203) & \\
\midrule\addlinespace[2.5pt]
Homeless Duration (days) & 834 (367) & 726 (415) & 0.28 \\ 
Network Size & 4.6 (4.4) & 4.8 (4.6) & -0.04 \\ 
Age (in years) & 45 (12) & 45 (12) & 0.02 \\ 
Age Groups &  &  & 0.10 \\ 
~~18-24 & 15 (2.4\%) & 6 (3.0\%) &  \\ 
~~25-34 & 121 (19\%) & 38 (19\%) &  \\ 
~~35-44 & 182 (29\%) & 59 (29\%) &  \\ 
~~45-54 & 160 (25\%) & 50 (25\%) &  \\ 
~~55-64 & 112 (18\%) & 41 (20\%) &  \\ 
~~65 or older & 39 (6.2\%) & 9 (4.4\%) &  \\ 
Gender &  &  & 0.05 \\ 
~~Male & 449 (71\%) & 141 (69\%) &  \\ 
~~Female & 171 (27\%) & 58 (29\%) &  \\ 
~~Non-binary & 9 (1.4\%) & 4 (2.0\%) &  \\ 
Race/Ethnicity &  &  & 0.25 \\ 
~~American Indian, Alaskan Native or Indigenous & 38 (6.0\%) & 8 (3.9\%) &  \\ 
~~Asian or Asian American & 11 (1.7\%) & 3 (1.5\%) &  \\ 
~~Black, African American, or African & 102 (16\%) & 20 (9.9\%) &  \\ 
~~Hispanic/Latino (any race) & 91 (14\%) & 28 (14\%) &  \\ 
~~Multiracial & 68 (11\%) & 20 (9.9\%) &  \\ 
~~Native Hawaiian or Pacific Islander & 10 (1.6\%) & 3 (1.5\%) &  \\ 
~~Non-Hispanic/Latino (race not reported) & 21 (3.3\%) & 9 (4.4\%) &  \\ 
~~White & 288 (46\%) & 112 (55\%) &  \\ 
Veteran Status & 70 (11\%) & 22 (11\%) & 0.01 \\ 
Housing Status &  &  & 0.10 \\ 
~~Permanent & 37 (5.9\%) & 16 (7.9\%) &  \\ 
~~Temporary & 86 (14\%) & 30 (15\%) &  \\ 
~~Sheltered & 61 (9.7\%) & 22 (11\%) &  \\ 
~~Unsheltered & 445 (71\%) & 135 (67\%) &  \\ 
Adverse Health Outcomes & & &\\
~~General Health & 324 (52\%) & 120 (59\%) & 0.15 \\ 
~~Substance Use & 289 (46\%) & 112 (56\%) & 0.20 \\ 
~~Mental Health & 224 (36\%) & 77 (38\%) & 0.04 \\ 
\bottomrule
\end{tabular*}
\begin{minipage}{\linewidth}
\textsuperscript{\textit{1}}Mean (SD); n (\%)\\
\textsuperscript{\textit{2}}Standardized Mean Difference\\
\end{minipage}
\end{table}

\subsection{Housing status and demographic characteristics}

Questions on housing status and demographic characteristics follow HUD guidelines, as the survey instrument was designed to replace the biennial HUD-mandated count of unsheltered people experiencing homelessness. Beyond ensuring certain standard measures (e.g., gender, race/ethnicity, age), it also provides comparability with Seattle/King County Continuum of Care data. Below, we discuss the details and data cleaning we performed for this analysis.

Respondents were asked where they slept the previous night to establish housing status. The original question provided 14 response options and an ``Other'' category for manually entered responses. To simplify, these responses were recoded into four categories: permanent housing, temporary housing, sheltered, and unsheltered. Low sample sizes in certain categories (e.g., workplace) required adjustments to enhance the reliability of analyses (see~\ref{variables}).

In addition to housing status, the survey collected information on the duration of respondents’ experience of homelessness. Participants specified how long they had experienced homelessness in terms of years, months, weeks, or days. We converted all of these into days and created a variable for the total number of days.

Demographic characteristics include numerical age\footnote{A propensity model with numerical age performed better at balancing than a categorical age.}, gender (Female, Male, and Non-binary),\footnote{The HUD standard gender measure includes five options; however, due to low sample size, we aggregate these into the non-binary category.} race and ethnicity, veteran status, and personal network size. Network size was measured using the close friends network, a list generator in which respondents were asked to list people they personally know who are unhoused or experiencing homelessness, excluding family living with them, along with each person's relation to the respondent (e.g., friend, family). Respondents could list up to 15 people, though this limit was not disclosed to them. For a detailed description of the network size measures and their properties, see \citet{almquist_understanding_2026}. Table~\ref{tbl:evict} displays the distribution of respondents by eviction status, demographic characteristics, and housing status. Our sample was predominantly middle-aged, with a higher proportion of male, White, and non-Hispanic/Latino-identified people.

Respondents who selected ``Do not know" in response to binary health items (e.g., mental health or substance use) were conservatively coded as “No.” This reflects a cautious interpretation of non-affirmative responses, on the grounds that such answers indicate uncertainty rather than clear identification with the condition. The primary exception is race and ethnicity, for which we constructed a combined race/ethnicity variable in accordance with the U.S. Office of Management and Budget \citep{office_of_management_and_budget_revisions_2024}. Responses marked as ``choose not to answer'' or skipped altogether were treated as missing and excluded from the analysis. Details of variable construction and coding procedures are available in the~\ref{variables}.

\subsection{Analytical Strategy}

To answer our research question, we aimed to estimate the direct effect of eviction on health outcomes, including self-rated general health, mental health, and substance use. Figure~\ref{fig:dag} presents a directed acyclic graph (DAG) summarizing the potential pathways between eviction and health. In this framework, housing status acts as a mediator, and demographic characteristics serve as confounders. The duration of homelessness also becomes an independent predictor of health outcomes. Based on the ``Background'' section, we account for the following relationships:

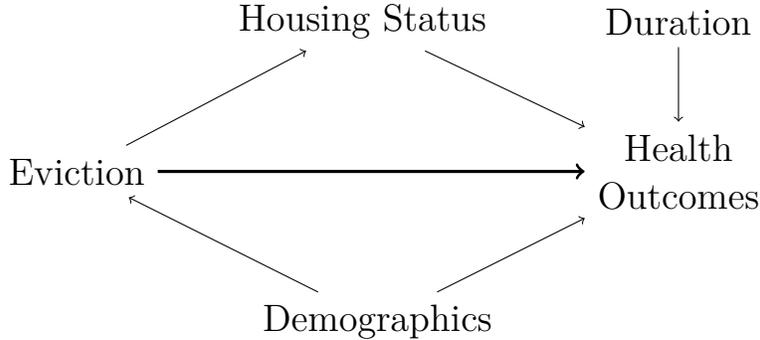
\begin{figure}
    \centering
    \large
    \begin{tikzpicture}
        \node (v0) at (0,-2) {Demographics};
        \node (v1) at (4,2) {Duration};
        \node (v2) at (-4,0) {Eviction};
        \node (v3)[align = center] at (4,0) {Health \\ Outcomes};
        \node (v4) at (-0.2,2) {Housing Status};
        \draw [->] (v0) edge (v2);
        \draw [->] (v0) edge (v3);
        \draw [->] (v1) edge (v3);
        \draw [very thick] [->] (v2) edge (v3);
        \draw [->] (v2) edge (v4);
        \draw [->] (v4) edge (v3);
    \end{tikzpicture}
    \caption{Directed Acyclic Graph (DAG) depicting relationships among variables.}
    \label{fig:dag}
\end{figure}

\begin{itemize}
    \item Eviction directly affects health outcomes. This is the primary effect we aim to estimate.
    \item The respondent's eviction history is closely linked to housing status. Individuals with a history of eviction are more likely to experience housing insecurity.
    \item Housing status directly affects health outcomes, with unsheltered individuals typically reporting poorer health.
    \item Longer durations of homelessness are associated with worse health outcomes.
    \item Demographic characteristics (age, gender, race, ethnicity, veteran status, and personal network size) are associated with both eviction risk and health outcomes. Eviction can also disrupt personal networks, which may in turn affect health.
\end{itemize}

These assumptions are supported by observed differences between our sample's evicted and non-evicted groups, particularly in housing status and racial/ethnic composition (see Table~\ref{tbl:evict}). To estimate the direct effect of eviction, we adjusted for all covariates with directed paths to the outcome, following the framework of \citet{chatton_g-computation_2020}.

Because confounders and mediators are present, standard logistic regression would conflate direct and indirect effects. We therefore employed a quasi-experimental strategy to isolate the direct effect of eviction on health outcomes.

\subsection{Inverse Probability of Treatment Weighting (IPTW)}

We employed an augmented version of inverse probability of treatment weighting (IPTW) consisting of two steps: (i) estimating a propensity model for eviction, and (ii) a weighted generalized linear model (GLM) for the health outcomes. Both models were estimated using maximum likelihood estimation (MLE).

As shown in Table~\ref{tbl:evict}, notable differences exist between the characteristics of the evicted and non-evicted samples. These imbalances reflect differential exposure to eviction across demographic groups and housing statuses. We used IPTW to create covariate-balanced groups to account for these differences, facilitating a quasi-experimental comparison.

First, we estimated propensity scores by fitting a logistic regression model with eviction as the dependent variable and covariates, including demographics, housing status, and duration of homelessness. Logistic regression provided better covariate balance compared to alternative specifications, such as probit models. We also evaluated newer balancing methods, including entropy balancing; however, given the marginal gains in performance, we selected logistic regression for the sake of parsimony. Covariate balance was assessed by examining standardized mean differences (SMDs) and empirical cumulative distribution functions before and after weighting. After weighting, all SMDs for covariates fell below 0.05, and all SMDs for squares, cubes, and two-way interactions fell below 0.1, indicating satisfactory balance (see~\ref{formal} for further details).

Next, we fit a weighted logistic regression model ( to estimate the effect of eviction on three binary outcomes: (i) self-rated health (low vs. high), (ii) serious mental health disorder (yes/no), and (iii) substance use disorder (yes/no). The primary exposure was eviction (yes/no), and we included interactions between eviction status and covariates. Inverse probability weights derived from the propensity model were incorporated to adjust for covariate imbalance. 

Average treatment effects (ATEs) were estimated using regression standardization (G-computation). Given that odds ratios are non-collapsible, we present marginal effects in the form of risk differences (RDs). Standard errors were calculated via the fractional weighted bootstrap (FWB) method, with 9,999 bootstrap replicates, refitting both the propensity and outcome models at each iteration. Although FWB provides more reliable estimates than traditional resampling methods, the differences are typically minor.

We used the same set of covariates for all three health outcomes for consistency and comparability across models. Our approach is often referred to as \textit{doubly robust}, meaning that consistent estimates of the average treatment effect can be obtained if either the propensity score model or the outcome regression model is correctly specified \citep{gabriel_inverse_2024, huntington-klein_effect_2021}. IPTW and weighted GLM were implemented using the \texttt{WeightIt} package \citep{greifer_weightit_2025}, and the G-computation step was carried out with the \texttt{MarginalEffects} package \citep{arel-bundock_how_2024}, both in \texttt{R}.

\section{Results}\label{sec:results}

Table~\ref{tbl:effect} presents the estimated effect of eviction in predicting self-rated health, substance use, and mental health, measured as risk differences. The effect of eviction on self-rated health is estimated at 0.083 (SE: 0.039, p-value: 0.03), indicating that eviction is associated with a 8.3 percentage point increase in the likelihood of reporting worse self-rated health. This result is statistically significant at 0.05 level, suggesting a meaningful association between eviction and poorer perceived health status. This finding aligns with previous research, highlighting the negative influence of housing instability on general health outcomes \citep{arcaya_effects_2013, martin_adults_2019}.

For substance use, the estimated effect of eviction is 0.095 (SE: 0.032, p-value: 0.003), indicating that individuals who experienced eviction are 9.5 percentage points more likely to report issues related to substance use compared to those who have not been evicted. This association is statistically significant at the 0.01 level, pointing to a meaningful relationship between eviction and increased substance use. The observed effect size is consistent with existing literature that links housing instability and forced displacements to higher rates of substance use \citep{mulia_economic_2014, murphy_housing_2013}. These findings underscore the potential role of eviction as a catalyst for substance use, possibly as a coping mechanism for the stress and instability associated with housing loss.

In contrast, the estimated effect of eviction on mental health is 0.012 (SE: 0.041, p-value: 0.8). This non-significant result suggests that eviction does not have a measurable independent effect on the self-reported mental health status of individuals in this sample. One possible explanation is that, since all individuals in our sample are experiencing homelessness, the mental health burden associated with homelessness itself may overshadow any additional effect attributable to eviction \citep{dunlap_behavioral_2016, pollack_health_2009}. The lack of a significant association could also reflect unmeasured confounders such as access to mental health services or social support networks. It is also possible that eviction undermines mental health in more indirect ways, mediated through changes in housing status or other stressors, rather than a direct effect.

Overall, the results indicate a significant and detrimental role of eviction in self-rated health and substance use, but no evidence of a direct effect on mental health outcomes. These findings highlight the importance of considering both direct and indirect pathways when examining the health consequences of eviction, as well as the need for targeted interventions to mitigate these adverse effects.

\vspace{12pt}
\begin{table}[!h]
\fontsize{12.0pt}{14.4pt}\selectfont
\caption{Effect of Eviction on Health Outcomes}
\centering
\begin{tabular*}{\linewidth}{@{\extracolsep{\fill}}lccc}
\toprule
  & General Health & Substance Use & Mental Health \\ 
\midrule\addlinespace[2.5pt]
Risk Difference & 0.083* & 0.095** & 0.012 \\ 
Std. Error & {(0.039)} & {(0.032)} & {(0.041)} \\ 
\midrule\addlinespace[2.5pt]
Num.Obs. & 831 & 823 & 822 \\ 
AIC & 2064.1 & 1882.8 & 1861.6 \\ 
BIC & 2234.1 & 2052.4 & 2031.2 \\ 
RMSE & 0.48 & 0.47 & 0.46 \\ 
\bottomrule
\end{tabular*}
\label{tbl:effect}
\raggedright
\begin{minipage}{\linewidth}
*\! p < 0.05, **\! p < 0.01, ***\! p < 0.001
\end{minipage}
\end{table}
The results presented in Table~\ref{tbl:effect} suggest eviction has a statistically significant and negative influence on self-rated health and substance use. The findings indicate that individuals who have been evicted are more likely to report poorer health and higher rates of substance use. However, no significant effect of eviction on mental health was observed in this analysis, suggesting the need for further research to explore potential mediating factors and unmeasured confounders that could influence this relationship.

\section{Discussion}\label{sec:discussion}

Our findings contribute to the growing body of literature highlighting the adverse health effects associated with eviction. Among people experiencing homelessness, we observed that the experience of eviction exacerbates negative health outcomes, including worse self-reported health and increased substance use. These results align with previous studies that document the severe health risks faced by the unsheltered population, who already experience higher rates of chronic disease, mental health issues, and premature mortality \citep{richards_unsheltered_2023, meyer_life_2023}. Despite the lower baseline health status typical among the unhoused, eviction remains a significant and independent predictor of further health deterioration, underscoring its profound harms for this population.

The association between eviction and health outcomes can be understood through several mechanisms. First, eviction imposes abrupt displacement, uprooting individuals from familiar environments and severing critical social networks. This sudden disruption can exacerbate stress and heighten vulnerabilities, potentially leading to increased substance use as a coping mechanism \citep{himmelstein_eviction_2021}. Additionally, the financial strain associated with eviction, including costs related to moving and legal fees, can intensify economic hardships, limiting access to healthcare and other essential resources. The emotional toll of eviction—marked by feelings of humiliation, loss of control, and isolation—may further compound the decline in self-rated health observed among evicted individuals \citep{hoke_health_2021, vasquez-vera_threat_2017}. Together, these acute pressures highlight eviction as a potent stressor that amplifies the adverse health effects of housing instability.

\subsection{Limitations}

While our study offers valuable insights into the relationship between eviction and health outcomes among people experiencing homelessness, it is not without limitations. The sample used in our analysis was recruited through RDS among the unsheltered population in King County, and is therefore primarily representative of that group. Although approximately 30\% of the sample includes sheltered, temporarily housed, and permanently housed individuals captured through the recruitment process (see Table~\ref{tbl:housing}), the survey was not designed to be representative of these subpopulations. As a result, the health profiles and eviction experiences of sheltered individuals may not be fully reflected in our findings.

Another potential limitation is the reliance on self-reported health measures. Although self-rated health is a widely accepted and validated indicator in public health research, it may differ from objective physiological assessments \citep{layes_whiners_2012}. Nonetheless, extensive research has demonstrated the strong predictive validity of self-rated health, including its ability to capture both physical and mental health dimensions \citep{goldberg_longitudinal_2001, garbarski_research_2016}. Future studies could benefit from integrating more precise health metrics, such as biomarkers or clinical evaluations, to provide a deeper understanding of the physiological effects of eviction.

\subsection{Policy Implications}

Unlike many structural inequalities that affect health and housing stability, eviction is a problem that can be directly addressed through policy interventions. During the COVID-19 pandemic, eviction moratoriums emerged as an important pandemic prevention policy \citep{benfer_eviction_2021}. Research suggests that these moratoriums were effective at reducing evictions, though racial and ethnic inequalities in eviction rates persisted \citep{leifheit_expiring_2021}. Moreover, the lifting of eviction moratoriums was associated with increased COVID-19 infection and death rates \citep{nande_effect_2021}, though questions about causality remain. These findings suggest that similar interventions warrant investigation in non-pandemic contexts, particularly given the health consequences documented in our study.

Our findings underscore the need for policies that treat eviction prevention as a public health priority. Several policy approaches have shown promise in preventing evictions and providing relief for households facing displacement. Legal interventions, including the right to counsel and just-cause eviction requirements, can protect tenants from unjust displacement \citep{rutan_concentrated_2021, desmond_evictions_2015}. Financial interventions—such as emergency rental assistance and support for basic needs like food, childcare, and transportation—can ease housing cost burdens before households fall behind on rent \citep{schapiro_effects_2022, kim_housing_2017}. Market-level interventions, including rent stabilization policies and support for tenant organizations, provide structural protections for precarious households struggling with rising rents and housing market pressures \citep{messamore_effect_2023}.

While these policy recommendations are well-established in housing and public health research, they merit continued emphasis given the cumulative evidence of evictions' harmful effects on health and well-being. These interventions are low-risk approaches that generate clear benefits, and while implementation challenges exist, the imperative remains prevention before households reach the point of being unable to pay rent. Addressing the root causes of housing instability requires increasing the production \citep{colburn_homelessness_2022} and preservation of affordable housing, expanding access to social support services, and addressing broader systemic inequalities that leave vulnerable populations at risk of both eviction and its associated health consequences.

\subsection{Conclusion}

Eviction destabilizes lives and disrupts essential social networks, making it difficult for individuals to secure alternative housing. This is particularly problematic for those with limited financial resources, unstable employment, or inadequate support systems. Consistent with prior research, our study finds that individuals who experience eviction are at a significantly higher risk of becoming homeless, as the eviction process often propels them into precarious living situations, including temporary shelters, overcrowded housing, or even sleeping on the streets. Addressing the structural drivers of eviction and implementing policies to prevent housing loss are crucial to improving the health and well-being of this highly vulnerable population.

\section*{Code availability}
The final code used to analyze our data will, upon acceptance, be made available in the Harvard Dataverse \url{TBD}.

\section*{Declaration of competing interest}

\noindent
Zack Almquist, Ihsan Kahveci, Tim Thomas, Nathalie Williams, Paul Hebert, and Amy Hagopian have no conflicts of interest to declare. Janelle Rothfolk and Cathea Carey, as employees of the King County Regional Homelessness Authority, acknowledge the importance of transparency and accountability in scientific research and peer review. The King County Regional Homelessness Authority employed Janelle Rothfolk during the 2022 Point-in-Time count and Janelle Rothfolk and Cathea Carey during the 2024 Point-in-Time. The King County Regional Homelessness Authority, in its role as the Community of Care (CoC) lead, both conducted and funded the King County 2022 and 2024 RDS surveys, which were carried out as a requirement of the HUD-mandated biannual 2022 and 2024 Point-in-Time counts. \textbf{Financial Interests}: Janelle Rothfolk and Cathea Carey declare that they have no financial interests, such as stocks, patents, or research grants, that may be perceived as affecting their objectivity in the peer review process. \textbf{Organizational Interests}: The King County Regional Homelessness Authority has a stake in addressing homelessness in King County and may be influenced by the outcomes of this survey project. As the Community of Care (CoC) lead, the King County Regional Homelessness Authority must complete a biannual unsheltered Point-in-Time count, as required by both Federal and State statutes, to maintain continued funding from the Department of Housing and Urban Development. As such, they have a stake in identifying valid methods for enumerating people experiencing unsheltered homelessness.

\section*{Data availability}

\noindent
HUD Data Exchange provides all official PIT totals from KCRHA (see
\url{https://www.hudexchange.info/resource/3031/pit-and-hic-data-since-2007/}). Individual data for 2022 and 2024 are available upon request from KCRHA and the University of Washington through Zack W. Almquist for the 2023 PIT data. Much of the 2023 data is available through the UW eScience Data Science for Social Good (DSSG) Program, \url{https://uwescience.github.io/DSSG2024_understanding_homelessness/}.

\section*{Acknowledgments}

\noindent
Partial support for this research came from an Eunice Kennedy Shriver National Institute of Child Health and Human Development research infrastructure grant, P2C HD042828, to the Center for Studies in Demography \& Ecology at the University of Washington; NSF CAREER Grant \#SES-2142964; UW Population Health Initiative Tier 3 Grant. The content is solely the authors' responsibility and does not necessarily represent the official NIH or NSF views. We also thank the UW eScience Data Science for Social Good (DSSG) Program students for their work cleaning up the 2023 RDS data. 
\clearpage

\bibliographystyle{elsarticle-harv} 
\bibliography{Eviction}

\begin{thebibliography}{8}
\providecommand{\natexlab}[1]{#1}
\providecommand{\url}[1]{\texttt{#1}}
\expandafter\ifx\csname urlstyle\endcsname\relax
  \providecommand{\doi}[1]{doi: #1}\else
  \providecommand{\doi}{doi: \begingroup \urlstyle{rm}\Url}\fi

\bibitem[Denman et~al.(2018)Denman, Baldwin, Betts, McQueen, and
  Tiro]{denman_reducing_2018}
Deanna~C. Denman, Austin~S. Baldwin, Andrea~C. Betts, Amy McQueen, and
  Jasmin~A. Tiro.
\newblock Reducing “{I} {Don}’t {Know}” {Responses} and {Missing}
  {Survey} {Data}: {Implications} for {Measurement}.
\newblock \emph{Medical decision making : an international journal of the
  Society for Medical Decision Making}, 38\penalty0 (6):\penalty0 673--682,
  August 2018.
\newblock ISSN 0272-989X.
\newblock \doi{10.1177/0272989X18785159}.
\newblock URL \url{https://www.ncbi.nlm.nih.gov/pmc/articles/PMC6076331/}.

\bibitem[Evans et~al.(2016)Evans, Sullivan, and Wallskog]{evans_impact_2016}
William~N. Evans, James~X. Sullivan, and Melanie Wallskog.
\newblock The impact of homelessness prevention programs on homelessness.
\newblock \emph{Science}, August 2016.
\newblock \doi{10.1126/science.aag0833}.
\newblock URL \url{https://www.science.org/doi/10.1126/science.aag0833}.

\bibitem[Gabriel et~al.(2024)Gabriel, Sachs, Martinussen, Waernbaum,
  Goetghebeur, Vansteelandt, and Sjölander]{gabriel_inverse_2024}
Erin~E. Gabriel, Michael~C. Sachs, Torben Martinussen, Ingeborg Waernbaum, Els
  Goetghebeur, Stijn Vansteelandt, and Arvid Sjölander.
\newblock Inverse probability of treatment weighting with generalized linear
  outcome models for doubly robust estimation.
\newblock \emph{Statistics in Medicine}, 43\penalty0 (3):\penalty0 534--547,
  2024.
\newblock ISSN 1097-0258.
\newblock \doi{10.1002/sim.9969}.
\newblock URL \url{https://onlinelibrary.wiley.com/doi/abs/10.1002/sim.9969}.
\newblock \_eprint: https://onlinelibrary.wiley.com/doi/pdf/10.1002/sim.9969.

\bibitem[García and Kim(2021)]{garcia_many_2021}
Ivis García and Keuntae Kim.
\newblock “{Many} of {Us} {Have} {Been} {Previously} {Evicted}”:
  {Exploring} the {Relationship} {Between} {Homelessness} and {Evictions}
  {Among} {Families} {Participating} in the {Rapid} {Rehousing} {Program} in
  {Salt} {Lake} {County}, {Utah}.
\newblock \emph{Housing Policy Debate}, September 2021.
\newblock ISSN 1051-1482.
\newblock URL
  \url{https://www.tandfonline.com/doi/abs/10.1080/10511482.2020.1828988}.

\bibitem[{Office of Management and
  Budget}(2024)]{office_of_management_and_budget_revisions_2024}
{Office of Management and Budget}.
\newblock \emph{Revisions to {OMB}'s {Statistical} {Policy} {Directive} {No}.
  15: {Standards} for {Maintaining}, {Collecting}, and {Presenting} {Federal}
  {Data} on {Race} and {Ethni} city}.
\newblock Federal Register, 89(61), 21868–21879, March 2024.

\bibitem[Passel(2024)]{passel_who_2024}
Mark Hugo~Lopez Passel, Jens Manuel Krogstad {and} Jeffrey~S.
\newblock Who is {Hispanic}?, September 2024.
\newblock URL
  \url{https://www.pewresearch.org/short-reads/2024/09/12/who-is-hispanic/}.

\bibitem[Tourangeau et~al.(2014)Tourangeau, Edwards, Johnson, Wolter, and
  Bates]{tourangeau_hard--survey_2014}
Roger Tourangeau, Brad Edwards, Timothy~P. Johnson, Kirk~M. Wolter, and Nancy
  Bates, editors.
\newblock \emph{Hard-to-{Survey} {Populations}}.
\newblock Cambridge University Press, Cambridge, 2014.
\newblock ISBN 978-1-107-03135-7.
\newblock \doi{10.1017/CBO9781139381635}.
\newblock URL
  \url{https://www.cambridge.org/core/product/F2910D9CDAF93E9E30DB1B3E6479F1E6}.

\bibitem[Waters et~al.(2022)Waters, Kiviniemi, Hay, and
  Orom]{waters_dismissing_2022}
Erika~A. Waters, Marc~T. Kiviniemi, Jennifer~L. Hay, and Heather Orom.
\newblock Dismissing “don’t know” responses to perceived risk survey
  items threatens the validity of theoretical and empirical behavior change
  research.
\newblock \emph{Perspectives on psychological science : a journal of the
  Association for Psychological Science}, 17\penalty0 (3):\penalty0 841--851,
  May 2022.
\newblock ISSN 1745-6916.
\newblock \doi{10.1177/17456916211017860}.
\newblock URL \url{https://www.ncbi.nlm.nih.gov/pmc/articles/PMC9081103/}.

\end{thebibliography}


\begin{thebibliography}{54}
\expandafter\ifx\csname natexlab\endcsname\relax\def\natexlab#1{#1}\fi
\providecommand{\url}[1]{\texttt{#1}}
\providecommand{\href}[2]{#2}
\providecommand{\path}[1]{#1}
\providecommand{\DOIprefix}{doi:}
\providecommand{\ArXivprefix}{arXiv:}
\providecommand{\URLprefix}{URL: }
\providecommand{\Pubmedprefix}{pmid:}
\providecommand{\doi}[1]{\href{http://dx.doi.org/#1}{\path{#1}}}
\providecommand{\Pubmed}[1]{\href{pmid:#1}{\path{#1}}}
\providecommand{\bibinfo}[2]{#2}
\ifx\xfnm\relax \def\xfnm[#1]{\unskip,\space#1}\fi
\bibitem[{Almquist et~al.(2024)Almquist, Kahveci, Hazel, Kajfasz, Rothfolk,
  Guilmette, Anderson, Ozeryansky and Hagopian}]{almquist_innovating_2024}
\bibinfo{author}{Almquist, Z.W.}, \bibinfo{author}{Kahveci, I.},
  \bibinfo{author}{Hazel, M.A.}, \bibinfo{author}{Kajfasz, O.},
  \bibinfo{author}{Rothfolk, J.}, \bibinfo{author}{Guilmette, C.},
  \bibinfo{author}{Anderson, M.C.}, \bibinfo{author}{Ozeryansky, L.},
  \bibinfo{author}{Hagopian, A.}, \bibinfo{year}{2024}.
\newblock \bibinfo{title}{Innovating a community-driven enumeration and needs
  assessment of people experiencing homelessness: a network sampling approach
  for the {HUD}-mandated point-in-time count}.
\newblock \bibinfo{journal}{American Journal of Epidemiology} ,
  \bibinfo{pages}{kwae342}\URLprefix
  \url{https://academic.oup.com/aje/advance-article/doi/10.1093/aje/kwae342/7749332},
  \DOIprefix\doi{10.1093/aje/kwae342}.
\bibitem[{Almquist et~al.(2026)Almquist, Kahveci, Kajfasz, Rothfolk and
  Hagopian}]{almquist_understanding_2026}
\bibinfo{author}{Almquist, Z.W.}, \bibinfo{author}{Kahveci, I.},
  \bibinfo{author}{Kajfasz, O.}, \bibinfo{author}{Rothfolk, J.},
  \bibinfo{author}{Hagopian, A.}, \bibinfo{year}{2026}.
\newblock \bibinfo{title}{Understanding the personal networks of people
  experiencing homelessness in {King} {County}, {WA} with aggregate relational
  data}.
\newblock \bibinfo{journal}{Social Networks} \bibinfo{volume}{86},
  \bibinfo{pages}{150--172}.
\newblock \URLprefix
  \url{https://linkinghub.elsevier.com/retrieve/pii/S0378873326000158},
  \DOIprefix\doi{10.1016/j.socnet.2026.02.003}.
\bibitem[{Arcaya et~al.(2013)Arcaya, Glymour, Chakrabarti, Christakis, Kawachi
  and Subramanian}]{arcaya_effects_2013}
\bibinfo{author}{Arcaya, M.}, \bibinfo{author}{Glymour, M.M.},
  \bibinfo{author}{Chakrabarti, P.}, \bibinfo{author}{Christakis, N.A.},
  \bibinfo{author}{Kawachi, I.}, \bibinfo{author}{Subramanian, S.V.},
  \bibinfo{year}{2013}.
\newblock \bibinfo{title}{Effects of proximate foreclosed properties on
  individuals' weight gain in {Massachusetts}, 1987-2008}.
\newblock \bibinfo{journal}{American Journal of Public Health}
  \bibinfo{volume}{103}, \bibinfo{pages}{e50--56}.
\newblock \DOIprefix\doi{10.2105/AJPH.2013.301460}.
\bibitem[{Arel-Bundock et~al.(2024)Arel-Bundock, Greifer and
  Heiss}]{arel-bundock_how_2024}
\bibinfo{author}{Arel-Bundock, V.}, \bibinfo{author}{Greifer, N.},
  \bibinfo{author}{Heiss, A.}, \bibinfo{year}{2024}.
\newblock \bibinfo{title}{How to {Interpret} {Statistical} {Models} {Using}
  marginaleffects for {R} and {Python}}.
\newblock \bibinfo{journal}{Journal of Statistical Software}
  \bibinfo{volume}{111}, \bibinfo{pages}{1--32}.
\newblock \URLprefix
  \url{https://www.jstatsoft.org/index.php/jss/article/view/v111i09},
  \DOIprefix\doi{10.18637/jss.v111.i09}.
\bibitem[{Benfer et~al.(2021)Benfer, Vlahov, Long, Walker-Wells, Pottenger,
  Gonsalves and Keene}]{benfer_eviction_2021}
\bibinfo{author}{Benfer, E.A.}, \bibinfo{author}{Vlahov, D.},
  \bibinfo{author}{Long, M.Y.}, \bibinfo{author}{Walker-Wells, E.},
  \bibinfo{author}{Pottenger, J.L.}, \bibinfo{author}{Gonsalves, G.},
  \bibinfo{author}{Keene, D.E.}, \bibinfo{year}{2021}.
\newblock \bibinfo{title}{Eviction, {Health} {Inequity}, and the {Spread} of
  {COVID}-19: {Housing} {Policy} as a {Primary} {Pandemic} {Mitigation}
  {Strategy}}.
\newblock \bibinfo{journal}{Journal of Urban Health} \bibinfo{volume}{98},
  \bibinfo{pages}{1--12}.
\newblock \URLprefix \url{https://doi.org/10.1007/s11524-020-00502-1},
  \DOIprefix\doi{10.1007/s11524-020-00502-1}.
\bibitem[{Burgard et~al.(2012)Burgard, Seefeldt and
  Zelner}]{burgard_housing_2012}
\bibinfo{author}{Burgard, S.A.}, \bibinfo{author}{Seefeldt, K.S.},
  \bibinfo{author}{Zelner, S.}, \bibinfo{year}{2012}.
\newblock \bibinfo{title}{Housing instability and health: {Findings} from the
  {Michigan} recession and recovery study}.
\newblock \bibinfo{journal}{Social Science \& Medicine} \bibinfo{volume}{75},
  \bibinfo{pages}{2215--2224}.
\newblock \URLprefix
  \url{https://www.sciencedirect.com/science/article/pii/S0277953612006272},
  \DOIprefix\doi{10.1016/j.socscimed.2012.08.020}.
\bibitem[{Cagney et~al.(2014)Cagney, Browning, Iveniuk and
  English}]{cagney_onset_2014}
\bibinfo{author}{Cagney, K.A.}, \bibinfo{author}{Browning, C.R.},
  \bibinfo{author}{Iveniuk, J.}, \bibinfo{author}{English, N.},
  \bibinfo{year}{2014}.
\newblock \bibinfo{title}{The onset of depression during the great recession:
  foreclosure and older adult mental health}.
\newblock \bibinfo{journal}{American Journal of Public Health}
  \bibinfo{volume}{104}, \bibinfo{pages}{498--505}.
\newblock \DOIprefix\doi{10.2105/AJPH.2013.301566}.
\bibitem[{Chatton et~al.(2020)Chatton, Le~Borgne, Leyrat, Gillaizeau, Rousseau,
  Barbin, Laplaud, Léger, Giraudeau and Foucher}]{chatton_g-computation_2020}
\bibinfo{author}{Chatton, A.}, \bibinfo{author}{Le~Borgne, F.},
  \bibinfo{author}{Leyrat, C.}, \bibinfo{author}{Gillaizeau, F.},
  \bibinfo{author}{Rousseau, C.}, \bibinfo{author}{Barbin, L.},
  \bibinfo{author}{Laplaud, D.}, \bibinfo{author}{Léger, M.},
  \bibinfo{author}{Giraudeau, B.}, \bibinfo{author}{Foucher, Y.},
  \bibinfo{year}{2020}.
\newblock \bibinfo{title}{G-computation, propensity score-based methods, and
  targeted maximum likelihood estimator for causal inference with different
  covariates sets: a comparative simulation study}.
\newblock \bibinfo{journal}{Scientific Reports} \bibinfo{volume}{10},
  \bibinfo{pages}{9219}.
\newblock \URLprefix \url{https://www.nature.com/articles/s41598-020-65917-x},
  \DOIprefix\doi{10.1038/s41598-020-65917-x}.
\bibitem[{Colburn and Aldern(2022)}]{colburn_homelessness_2022}
\bibinfo{author}{Colburn, G.}, \bibinfo{author}{Aldern, C.},
  \bibinfo{year}{2022}.
\newblock \bibinfo{title}{Homelessness {Is} a {Housing} {Problem}: {How}
  {Structural} {Factors} {Explain} {U}.{S}. {Patterns}}.
\newblock \bibinfo{publisher}{University of California Press}.
\newblock \URLprefix \url{https://books.google.com/books?id=guxcEAAAQBAJ}.
\bibitem[{Desmond(2012)}]{desmond_eviction_2012}
\bibinfo{author}{Desmond, M.}, \bibinfo{year}{2012}.
\newblock \bibinfo{title}{Eviction and the {Reproduction} of {Urban}
  {Poverty}}.
\newblock \bibinfo{journal}{American Journal of Sociology}
  \bibinfo{volume}{118}, \bibinfo{pages}{88--133}.
\newblock \URLprefix \url{https://doi.org/10.1086/666082},
  \DOIprefix\doi{10.1086/666082}.
\bibitem[{Desmond and Kimbro(2015)}]{desmond_evictions_2015}
\bibinfo{author}{Desmond, M.}, \bibinfo{author}{Kimbro, R.T.},
  \bibinfo{year}{2015}.
\newblock \bibinfo{title}{Eviction's {Fallout}: {Housing}, {Hardship}, and
  {Health}}.
\newblock \bibinfo{journal}{Social Forces} \bibinfo{volume}{94},
  \bibinfo{pages}{295--324}.
\newblock \URLprefix
  \url{https://academic.oup.com/sf/article-lookup/doi/10.1093/sf/sov044},
  \DOIprefix\doi{10.1093/sf/sov044}.
\bibitem[{Dunlap et~al.(2016)Dunlap, Han, Dowd, Cowell, Forman-Hoffman, Davies
  and Colpe}]{dunlap_behavioral_2016}
\bibinfo{author}{Dunlap, L.J.}, \bibinfo{author}{Han, B.},
  \bibinfo{author}{Dowd, W.N.}, \bibinfo{author}{Cowell, A.J.},
  \bibinfo{author}{Forman-Hoffman, V.L.}, \bibinfo{author}{Davies, M.C.},
  \bibinfo{author}{Colpe, L.J.}, \bibinfo{year}{2016}.
\newblock \bibinfo{title}{Behavioral {Health} {Outcomes} {Among} {Adults}:
  {Associations} {With} {Individual} and {Community}-{Level} {Economic}
  {Conditions}}.
\newblock \bibinfo{journal}{Psychiatric Services (Washington, D.C.)}
  \bibinfo{volume}{67}, \bibinfo{pages}{71--77}.
\newblock \DOIprefix\doi{10.1176/appi.ps.201400016}.
\bibitem[{Fowle and Fyall(2024)}]{fowle_evading_2024}
\bibinfo{author}{Fowle, M.}, \bibinfo{author}{Fyall, R.}, \bibinfo{year}{2024}.
\newblock \bibinfo{title}{Evading the eviction moratorium: {Changing} patterns
  in formal and informal evictions and eviction tactics during the {COVID}-19
  pandemic}.
\newblock \bibinfo{journal}{Journal of Urban Affairs} ,
  \bibinfo{pages}{1--19}\URLprefix
  \url{https://www.tandfonline.com/doi/full/10.1080/07352166.2024.2415937},
  \DOIprefix\doi{10.1080/07352166.2024.2415937}.
\bibitem[{Gabriel et~al.(2024)Gabriel, Sachs, Martinussen, Waernbaum,
  Goetghebeur, Vansteelandt and Sjölander}]{gabriel_inverse_2024}
\bibinfo{author}{Gabriel, E.E.}, \bibinfo{author}{Sachs, M.C.},
  \bibinfo{author}{Martinussen, T.}, \bibinfo{author}{Waernbaum, I.},
  \bibinfo{author}{Goetghebeur, E.}, \bibinfo{author}{Vansteelandt, S.},
  \bibinfo{author}{Sjölander, A.}, \bibinfo{year}{2024}.
\newblock \bibinfo{title}{Inverse probability of treatment weighting with
  generalized linear outcome models for doubly robust estimation}.
\newblock \bibinfo{journal}{Statistics in Medicine} \bibinfo{volume}{43},
  \bibinfo{pages}{534--547}.
\newblock \URLprefix
  \url{https://onlinelibrary.wiley.com/doi/abs/10.1002/sim.9969},
  \DOIprefix\doi{10.1002/sim.9969}. \bibinfo{note}{\_eprint:
  https://onlinelibrary.wiley.com/doi/pdf/10.1002/sim.9969}.
\bibitem[{Garbarski(2016)}]{garbarski_research_2016}
\bibinfo{author}{Garbarski, D.}, \bibinfo{year}{2016}.
\newblock \bibinfo{title}{Research in and {Prospects} for the {Measurement} of
  {Health} {Using} {Self}-{Rated} {Health}}.
\newblock \bibinfo{journal}{Public Opinion Quarterly} \bibinfo{volume}{80},
  \bibinfo{pages}{977--997}.
\newblock \URLprefix \url{https://doi.org/10.1093/poq/nfw033},
  \DOIprefix\doi{10.1093/poq/nfw033}.
\bibitem[{Garcia et~al.(2024)Garcia, Doran and
  Kushel}]{garcia_homelessness_2024}
\bibinfo{author}{Garcia, C.}, \bibinfo{author}{Doran, K.},
  \bibinfo{author}{Kushel, M.}, \bibinfo{year}{2024}.
\newblock \bibinfo{title}{Homelessness {And} {Health}: {Factors}, {Evidence},
  {Innovations} {That} {Work}, {And} {Policy} {Recommendations}: {An} overview
  of factors and policy recommendations pertaining to homelessness and health}.
\newblock \bibinfo{journal}{Health Affairs} \bibinfo{volume}{43},
  \bibinfo{pages}{164--171}.
\newblock \URLprefix
  \url{http://www.healthaffairs.org/doi/10.1377/hlthaff.2023.01049},
  \DOIprefix\doi{10.1377/hlthaff.2023.01049}.
\bibitem[{García and Kim(2021)}]{garcia_many_2021}
\bibinfo{author}{García, I.}, \bibinfo{author}{Kim, K.}, \bibinfo{year}{2021}.
\newblock \bibinfo{title}{“{Many} of {Us} {Have} {Been} {Previously}
  {Evicted}”: {Exploring} the {Relationship} {Between} {Homelessness} and
  {Evictions} {Among} {Families} {Participating} in the {Rapid} {Rehousing}
  {Program} in {Salt} {Lake} {County}, {Utah}}.
\newblock \bibinfo{journal}{Housing Policy Debate} \URLprefix
  \url{https://www.tandfonline.com/doi/abs/10.1080/10511482.2020.1828988}.
\bibitem[{Goldberg et~al.(2001)Goldberg, Gueguen, Schmaus, Nakache and
  Goldberg}]{goldberg_longitudinal_2001}
\bibinfo{author}{Goldberg, P.}, \bibinfo{author}{Gueguen, A.},
  \bibinfo{author}{Schmaus, A.}, \bibinfo{author}{Nakache, J.},
  \bibinfo{author}{Goldberg, M.}, \bibinfo{year}{2001}.
\newblock \bibinfo{title}{Longitudinal study of associations between perceived
  health status and self reported diseases in the {French} {Gazel} cohort}.
\newblock \bibinfo{journal}{Journal of Epidemiology and Community Health}
  \bibinfo{volume}{55}, \bibinfo{pages}{233--238}.
\newblock \URLprefix
  \url{https://www.ncbi.nlm.nih.gov/pmc/articles/PMC1731872/},
  \DOIprefix\doi{10.1136/jech.55.4.233}.
\bibitem[{Graetz et~al.(2024)Graetz, Gershenson, Porter, Sandler, Lemmerman and
  Desmond}]{graetz_impacts_2024}
\bibinfo{author}{Graetz, N.}, \bibinfo{author}{Gershenson, C.},
  \bibinfo{author}{Porter, S.R.}, \bibinfo{author}{Sandler, D.H.},
  \bibinfo{author}{Lemmerman, E.}, \bibinfo{author}{Desmond, M.},
  \bibinfo{year}{2024}.
\newblock \bibinfo{title}{The impacts of rent burden and eviction on mortality
  in the {United} {States}, 2000–2019}.
\newblock \bibinfo{journal}{Social Science \& Medicine} \bibinfo{volume}{340},
  \bibinfo{pages}{116398}.
\newblock \URLprefix
  \url{https://www.sciencedirect.com/science/article/pii/S0277953623007554},
  \DOIprefix\doi{10.1016/j.socscimed.2023.116398}.
\bibitem[{Grainger(2021)}]{grainger_its_2021}
\bibinfo{author}{Grainger, G.L.}, \bibinfo{year}{2021}.
\newblock \bibinfo{title}{“{It}'s {Not} a {Pattern} of {Behavior}”: {Proxy}
  {Deflection} of {Eviction} {Stigma} by {Community} {Care} {Providers}}.
\newblock \bibinfo{journal}{Symbolic Interaction} \bibinfo{volume}{44},
  \bibinfo{pages}{479--503}.
\newblock \URLprefix
  \url{https://onlinelibrary.wiley.com/doi/abs/10.1002/symb.511},
  \DOIprefix\doi{10.1002/symb.511}. \bibinfo{note}{\_eprint:
  https://onlinelibrary.wiley.com/doi/pdf/10.1002/symb.511}.
\bibitem[{Greifer(2025)}]{greifer_weightit_2025}
\bibinfo{author}{Greifer, N.}, \bibinfo{year}{2025}.
\newblock \bibinfo{title}{{WeightIt}: {Weighting} for {Covariate} {Balance} in
  {Observational} {Studies}}.
\newblock \URLprefix \url{https://ngreifer.github.io/WeightIt/}.
\bibitem[{Gromis and Desmond(2021)}]{gromis_estimating_2021}
\bibinfo{author}{Gromis, A.}, \bibinfo{author}{Desmond, M.},
  \bibinfo{year}{2021}.
\newblock \bibinfo{title}{Estimating the prevalence of eviction in the {United}
  {States}}.
\newblock \bibinfo{journal}{Cityscape} \bibinfo{volume}{23},
  \bibinfo{pages}{279--290}.
\bibitem[{Gromis et~al.(2022)Gromis, Fellows, Hendrickson, Edmonds, Leung,
  Porton and Desmond}]{gromis_estimating_2022}
\bibinfo{author}{Gromis, A.}, \bibinfo{author}{Fellows, I.},
  \bibinfo{author}{Hendrickson, J.R.}, \bibinfo{author}{Edmonds, L.},
  \bibinfo{author}{Leung, L.}, \bibinfo{author}{Porton, A.},
  \bibinfo{author}{Desmond, M.}, \bibinfo{year}{2022}.
\newblock \bibinfo{title}{Estimating eviction prevalence across the {United}
  {States}}.
\newblock \bibinfo{journal}{Proceedings of the National Academy of Sciences}
  \bibinfo{volume}{119}, \bibinfo{pages}{e2116169119}.
\newblock \URLprefix \url{https://pnas.org/doi/full/10.1073/pnas.2116169119},
  \DOIprefix\doi{10.1073/pnas.2116169119}.
\bibitem[{Heckathorn(1997)}]{heckathorn_respondent-driven_1997}
\bibinfo{author}{Heckathorn, D.D.}, \bibinfo{year}{1997}.
\newblock \bibinfo{title}{Respondent-driven sampling: a new approach to the
  study of hidden populations}.
\newblock \bibinfo{journal}{Social problems} \bibinfo{volume}{44},
  \bibinfo{pages}{174--199}.
\bibitem[{Himmelstein and Desmond(2021)}]{himmelstein_eviction_2021}
\bibinfo{author}{Himmelstein, G.}, \bibinfo{author}{Desmond, M.},
  \bibinfo{year}{2021}.
\newblock \bibinfo{title}{Eviction {And} {Health}: {A} {Vicious} {Cycle}
  {Exacerbated} {By} {A} {Pandemic}}.
\newblock \bibinfo{type}{Technical Report}. Project HOPE.
\newblock \URLprefix
  \url{https://www.healthaffairs.org/do/10.1377/hpb20210315.747908/full/},
  \DOIprefix\doi{10.1377/hpb20210315.747908}.
\bibitem[{Hoke and Boen(2021)}]{hoke_health_2021}
\bibinfo{author}{Hoke, M.K.}, \bibinfo{author}{Boen, C.E.},
  \bibinfo{year}{2021}.
\newblock \bibinfo{title}{The health impacts of eviction: {Evidence} from the
  national longitudinal study of adolescent to adult health}.
\newblock \bibinfo{journal}{Social Science \& Medicine} \bibinfo{volume}{273},
  \bibinfo{pages}{113742}.
\newblock \URLprefix
  \url{https://www.sciencedirect.com/science/article/pii/S0277953621000745},
  \DOIprefix\doi{10.1016/j.socscimed.2021.113742}.
\bibitem[{Huntington-Klein(2021)}]{huntington-klein_effect_2021}
\bibinfo{author}{Huntington-Klein, N.}, \bibinfo{year}{2021}.
\newblock \bibinfo{title}{The {Effect}: {An} {Introduction} to {Research}
  {Design} and {Causality}}.
\newblock \bibinfo{publisher}{CRC Press}.
\newblock \URLprefix \url{https://books.google.com/books?id=f0NOEAAAQBAJ}.
\bibitem[{Idler and Benyamini(1997)}]{idler_self-rated_1997}
\bibinfo{author}{Idler, E.L.}, \bibinfo{author}{Benyamini, Y.},
  \bibinfo{year}{1997}.
\newblock \bibinfo{title}{Self-rated health and mortality: a review of
  twenty-seven community studies}.
\newblock \bibinfo{journal}{Journal of Health and Social Behavior}
  \bibinfo{volume}{38}, \bibinfo{pages}{21--37}.
\bibitem[{Jylhä(2009)}]{jylha_what_2009}
\bibinfo{author}{Jylhä, M.}, \bibinfo{year}{2009}.
\newblock \bibinfo{title}{What is self-rated health and why does it predict
  mortality? {Towards} a unified conceptual model}.
\newblock \bibinfo{journal}{Social Science \& Medicine (1982)}
  \bibinfo{volume}{69}, \bibinfo{pages}{307--316}.
\newblock \DOIprefix\doi{10.1016/j.socscimed.2009.05.013}.
\bibitem[{Kim et~al.(2017)Kim, Burgard and Seefeldt}]{kim_housing_2017}
\bibinfo{author}{Kim, H.}, \bibinfo{author}{Burgard, S.A.},
  \bibinfo{author}{Seefeldt, K.S.}, \bibinfo{year}{2017}.
\newblock \bibinfo{title}{Housing {Assistance} and {Housing} {Insecurity}: {A}
  {Study} of {Renters} in {Southeastern} {Michigan} in the {Wake} of the
  {Great} {Recession}}.
\newblock \bibinfo{journal}{Social Service Review} \bibinfo{volume}{91},
  \bibinfo{pages}{41--70}.
\newblock \URLprefix
  \url{https://www.journals.uchicago.edu/doi/10.1086/690681},
  \DOIprefix\doi{10.1086/690681}.
\bibitem[{{King County Reginonal Homelessness
  Authority}(2025)}]{kcrha_pit_2025}
\bibinfo{author}{{King County Reginonal Homelessness Authority}},
  \bibinfo{year}{2025}.
\newblock \bibinfo{title}{King {County} 2024 {Point} {In} {Time} {Count}}.
\newblock \bibinfo{type}{Technical Report}. \bibinfo{address}{Seattle, WA}.
\newblock \URLprefix
  \url{https://kcrha.org/wp-content/uploads/2025/05/Point-in-Time-Count-2024_King-County_final.pdf}.
\bibitem[{Kushel and Moore(2023)}]{kushel_toward_2023}
\bibinfo{author}{Kushel, M.}, \bibinfo{author}{Moore, T.},
  \bibinfo{year}{2023}.
\newblock \bibinfo{title}{Toward a {New} {Understanding}: {The} {California}
  {Statewide} {Study} of {People} {Experiencing} {Homelessness}}.
\newblock \bibinfo{type}{Technical Report}. UCSF Benioff Homelessness and
  Housing Initiative.
\newblock \URLprefix
  \url{https://homelessness.ucsf.edu/resources/reports/toward-new-understanding-california-statewide-study-people-experiencing}.
\bibitem[{Layes et~al.(2012)Layes, Asada and Kephart}]{layes_whiners_2012}
\bibinfo{author}{Layes, A.}, \bibinfo{author}{Asada, Y.},
  \bibinfo{author}{Kephart, G.}, \bibinfo{year}{2012}.
\newblock \bibinfo{title}{Whiners and deniers – {What} does self-rated health
  measure?}
\newblock \bibinfo{journal}{Social Science \& Medicine} \bibinfo{volume}{75},
  \bibinfo{pages}{1--9}.
\newblock \URLprefix
  \url{https://www.sciencedirect.com/science/article/pii/S0277953611006745},
  \DOIprefix\doi{10.1016/j.socscimed.2011.10.030}.
\bibitem[{Leifheit et~al.(2021)Leifheit, Linton, Raifman, Schwartz, Benfer,
  Zimmerman and Pollack}]{leifheit_expiring_2021}
\bibinfo{author}{Leifheit, K.M.}, \bibinfo{author}{Linton, S.L.},
  \bibinfo{author}{Raifman, J.}, \bibinfo{author}{Schwartz, G.L.},
  \bibinfo{author}{Benfer, E.A.}, \bibinfo{author}{Zimmerman, F.J.},
  \bibinfo{author}{Pollack, C.E.}, \bibinfo{year}{2021}.
\newblock \bibinfo{title}{Expiring {Eviction} {Moratoriums} and {COVID}-19
  {Incidence} and {Mortality}}.
\newblock \bibinfo{journal}{American Journal of Epidemiology}
  \bibinfo{volume}{190}, \bibinfo{pages}{2503--2510}.
\newblock \URLprefix
  \url{https://academic.oup.com/aje/article/190/12/2503/6328194},
  \DOIprefix\doi{10.1093/aje/kwab196}.
\bibitem[{Martin et~al.(2019)Martin, Liaw, Bazemore, Jetty, Petterson and
  Kushel}]{martin_adults_2019}
\bibinfo{author}{Martin, P.}, \bibinfo{author}{Liaw, W.},
  \bibinfo{author}{Bazemore, A.}, \bibinfo{author}{Jetty, A.},
  \bibinfo{author}{Petterson, S.}, \bibinfo{author}{Kushel, M.},
  \bibinfo{year}{2019}.
\newblock \bibinfo{title}{Adults with {Housing} {Insecurity} {Have} {Worse}
  {Access} to {Primary} and {Preventive} {Care}}.
\newblock \bibinfo{journal}{Journal of the American Board of Family Medicine:
  JABFM} \bibinfo{volume}{32}, \bibinfo{pages}{521--530}.
\newblock \DOIprefix\doi{10.3122/jabfm.2019.04.180374}.
\bibitem[{Messamore(2023)}]{messamore_effect_2023}
\bibinfo{author}{Messamore, A.}, \bibinfo{year}{2023}.
\newblock \bibinfo{title}{The {Effect} of {Community} {Organizing} on
  {Landlords}’ {Use} of {Eviction} {Filing}: {Evidence} from {U}.{S}.
  {Cities}}.
\newblock \bibinfo{journal}{Social Problems} \bibinfo{volume}{70},
  \bibinfo{pages}{809--830}.
\newblock \URLprefix \url{https://doi.org/10.1093/socpro/spac061},
  \DOIprefix\doi{10.1093/socpro/spac061}.
\bibitem[{Meyer et~al.(2023)Meyer, Wyse and Logani}]{meyer_life_2023}
\bibinfo{author}{Meyer, B.D.}, \bibinfo{author}{Wyse, A.},
  \bibinfo{author}{Logani, I.}, \bibinfo{year}{2023}.
\newblock \bibinfo{title}{Life and {Death} at the {Margins} of {Society}: {The}
  {Mortality} of the {U}.{S}. {Homeless} {Population}}.
\newblock Number \bibinfo{number}{no. w31843} in \bibinfo{series}{{NBER}
  working paper series}, \bibinfo{publisher}{National Bureau of Economic
  Research}, \bibinfo{address}{Cambridge, Mass}.
\bibitem[{Mulia et~al.(2014)Mulia, Zemore, Murphy, Liu and
  Catalano}]{mulia_economic_2014}
\bibinfo{author}{Mulia, N.}, \bibinfo{author}{Zemore, S.E.},
  \bibinfo{author}{Murphy, R.}, \bibinfo{author}{Liu, H.},
  \bibinfo{author}{Catalano, R.}, \bibinfo{year}{2014}.
\newblock \bibinfo{title}{Economic loss and alcohol consumption and problems
  during the 2008 to 2009 {U}.{S}. recession}.
\newblock \bibinfo{journal}{Alcoholism, Clinical and Experimental Research}
  \bibinfo{volume}{38}, \bibinfo{pages}{1026--1034}.
\newblock \DOIprefix\doi{10.1111/acer.12301}.
\bibitem[{Murphy et~al.(2013)Murphy, Zemore and Mulia}]{murphy_housing_2013}
\bibinfo{author}{Murphy, R.D.}, \bibinfo{author}{Zemore, S.E.},
  \bibinfo{author}{Mulia, N.}, \bibinfo{year}{2013}.
\newblock \bibinfo{title}{Housing {Instability} and {Alcohol} {Problems} during
  the 2007–2009 {US} {Recession}: the {Moderating} {Role} of {Perceived}
  {Family} {Support}}.
\newblock \bibinfo{journal}{Journal of Urban Health : Bulletin of the New York
  Academy of Medicine} \bibinfo{volume}{91}, \bibinfo{pages}{17}.
\newblock \URLprefix \url{https://pmc.ncbi.nlm.nih.gov/articles/PMC3907617/},
  \DOIprefix\doi{10.1007/s11524-013-9813-z}.
\bibitem[{Nande et~al.(2021)Nande, Sheen, Walters, Klein, Chinazzi, Gheorghe,
  Adlam, Shinnick, Tejeda, Scarpino, Vespignani, Greenlee, Schneider, Levy and
  Hill}]{nande_effect_2021}
\bibinfo{author}{Nande, A.}, \bibinfo{author}{Sheen, J.},
  \bibinfo{author}{Walters, E.L.}, \bibinfo{author}{Klein, B.},
  \bibinfo{author}{Chinazzi, M.}, \bibinfo{author}{Gheorghe, A.H.},
  \bibinfo{author}{Adlam, B.}, \bibinfo{author}{Shinnick, J.},
  \bibinfo{author}{Tejeda, M.F.}, \bibinfo{author}{Scarpino, S.V.},
  \bibinfo{author}{Vespignani, A.}, \bibinfo{author}{Greenlee, A.J.},
  \bibinfo{author}{Schneider, D.}, \bibinfo{author}{Levy, M.Z.},
  \bibinfo{author}{Hill, A.L.}, \bibinfo{year}{2021}.
\newblock \bibinfo{title}{The effect of eviction moratoria on the transmission
  of {SARS}-{CoV}-2}.
\newblock \bibinfo{journal}{Nature Communications} \bibinfo{volume}{12},
  \bibinfo{pages}{2274}.
\newblock \URLprefix \url{https://www.nature.com/articles/s41467-021-22521-5},
  \DOIprefix\doi{10.1038/s41467-021-22521-5}.
\bibitem[{Niccolai et~al.(2019)Niccolai, Blankenship and
  Keene}]{niccolai_eviction_2019}
\bibinfo{author}{Niccolai, L.M.}, \bibinfo{author}{Blankenship, K.M.},
  \bibinfo{author}{Keene, D.E.}, \bibinfo{year}{2019}.
\newblock \bibinfo{title}{Eviction {From} {Renter}-occupied {Households} and
  {Rates} of {Sexually} {Transmitted} {Infections}: {A} {County}-level
  {Ecological} {Analysis}}.
\newblock \bibinfo{journal}{Sexually Transmitted Diseases}
  \bibinfo{volume}{46}, \bibinfo{pages}{63--68}.
\newblock \DOIprefix\doi{10.1097/OLQ.0000000000000904}.
\bibitem[{{Office of Management and
  Budget}(2024)}]{office_of_management_and_budget_revisions_2024}
\bibinfo{author}{{Office of Management and Budget}}, \bibinfo{year}{2024}.
\newblock \bibinfo{title}{Revisions to {OMB}'s {Statistical} {Policy}
  {Directive} {No}. 15: {Standards} for {Maintaining}, {Collecting}, and
  {Presenting} {Federal} {Data} on {Race} and {Ethni} city}.
\newblock \bibinfo{publisher}{Federal Register, 89(61), 21868–21879}.
\bibitem[{Pollack and Lynch(2009)}]{pollack_health_2009}
\bibinfo{author}{Pollack, C.E.}, \bibinfo{author}{Lynch, J.},
  \bibinfo{year}{2009}.
\newblock \bibinfo{title}{Health {Status} of {People} {Undergoing}
  {Foreclosure} in the {Philadelphia} {Region}}.
\newblock \bibinfo{journal}{American Journal of Public Health}
  \bibinfo{volume}{99}, \bibinfo{pages}{1833}.
\newblock \URLprefix \url{https://pmc.ncbi.nlm.nih.gov/articles/PMC2741520/},
  \DOIprefix\doi{10.2105/AJPH.2009.161380}.
\bibitem[{Richards and Kuhn(2023)}]{richards_unsheltered_2023}
\bibinfo{author}{Richards, J.}, \bibinfo{author}{Kuhn, R.},
  \bibinfo{year}{2023}.
\newblock \bibinfo{title}{Unsheltered {Homelessness} and {Health}: {A}
  {Literature} {Review}}.
\newblock \bibinfo{journal}{AJPM Focus} \bibinfo{volume}{2},
  \bibinfo{pages}{100043}.
\newblock \URLprefix
  \url{https://www.sciencedirect.com/science/article/pii/S2773065422000414},
  \DOIprefix\doi{10.1016/j.focus.2022.100043}.
\bibitem[{Rutan and Desmond(2021)}]{rutan_concentrated_2021}
\bibinfo{author}{Rutan, D.Q.}, \bibinfo{author}{Desmond, M.},
  \bibinfo{year}{2021}.
\newblock \bibinfo{title}{The {Concentrated} {Geography} of {Eviction}}.
\newblock \bibinfo{journal}{The ANNALS of the American Academy of Political and
  Social Science} \bibinfo{volume}{693}, \bibinfo{pages}{64--81}.
\newblock \URLprefix \url{https://doi.org/10.1177/0002716221991458},
  \DOIprefix\doi{10.1177/0002716221991458}.
\bibitem[{Schapiro et~al.(2022)Schapiro, Blankenship, Rosenberg and
  Keene}]{schapiro_effects_2022}
\bibinfo{author}{Schapiro, R.}, \bibinfo{author}{Blankenship, K.},
  \bibinfo{author}{Rosenberg, A.}, \bibinfo{author}{Keene, D.},
  \bibinfo{year}{2022}.
\newblock \bibinfo{title}{The {Effects} of {Rental} {Assistance} on {Housing}
  {Stability}, {Quality}, {Autonomy}, and {Affordability}}.
\newblock \bibinfo{journal}{Housing policy debate} \bibinfo{volume}{32},
  \bibinfo{pages}{456--472}.
\newblock \URLprefix \url{https://pmc.ncbi.nlm.nih.gov/articles/PMC9173361/},
  \DOIprefix\doi{10.1080/10511482.2020.1846067}.
\bibitem[{Thomas and Schwinghammer(2024)}]{thomas_washington_2024}
\bibinfo{author}{Thomas, T.}, \bibinfo{author}{Schwinghammer, M.},
  \bibinfo{year}{2024}.
\newblock \bibinfo{title}{Washington {State} {Eviction} {Filings}}.
\newblock \URLprefix \url{https://evictionresearch.net/washington/}.
\bibitem[{Tourangeau et~al.(2014)Tourangeau, Edwards, Johnson, Wolter and
  Bates}]{tourangeau_hard--survey_2014}
\bibinfo{editor}{Tourangeau, R.}, \bibinfo{editor}{Edwards, B.},
  \bibinfo{editor}{Johnson, T.P.}, \bibinfo{editor}{Wolter, K.M.},
  \bibinfo{editor}{Bates, N.} (Eds.), \bibinfo{year}{2014}.
\newblock \bibinfo{title}{Hard-to-{Survey} {Populations}}.
\newblock \bibinfo{publisher}{Cambridge University Press},
  \bibinfo{address}{Cambridge}.
\newblock \URLprefix
  \url{https://www.cambridge.org/core/product/F2910D9CDAF93E9E30DB1B3E6479F1E6},
  \DOIprefix\doi{10.1017/CBO9781139381635}.
\bibitem[{Tsai and Alarcón(2022)}]{tsai_annual_2022}
\bibinfo{author}{Tsai, J.}, \bibinfo{author}{Alarcón, J.},
  \bibinfo{year}{2022}.
\newblock \bibinfo{title}{The {Annual} {Homeless} {Point}-in-{Time} {Count}:
  {Limitations} and {Two} {Different} {Solutions}}.
\newblock \bibinfo{journal}{American Journal of Public Health}
  \bibinfo{volume}{112}, \bibinfo{pages}{633--637}.
\newblock \URLprefix
  \url{https://ajph.aphapublications.org/doi/full/10.2105/AJPH.2021.306640},
  \DOIprefix\doi{10.2105/AJPH.2021.306640}.
\bibitem[{{U.S. Department of Housing and Urban
  Development}()}]{us_department_of_housing_and_urban_development_hmis_nodate}
\bibinfo{author}{{U.S. Department of Housing and Urban Development}}, .
\newblock \bibinfo{title}{{HMIS} data standards: {Common} program data
  elements: 4.09 mental health disorder}.
\newblock \URLprefix
  \url{https://www.hudexchange.info/programs/hmis/hmis-data-standards/standards/common-program-specific-data-elements/409-mental-health-disorder/}.
\bibitem[{{U.S. Department of Housing and Urban
  Development}(2011)}]{us_department_of_housing_and_urban_development_homeless_2011}
\bibinfo{author}{{U.S. Department of Housing and Urban Development}},
  \bibinfo{year}{2011}.
\newblock \bibinfo{title}{Homeless {Definition} and {Recordkeeping}
  {Requirements}}.
\newblock \URLprefix
  \url{https://www.hudexchange.info/resources/documents/HomelessDefinition_RecordkeepingRequirementsandCriteria.pdf}.
\bibitem[{Volz and Heckathorn(2008)}]{volz_probability_2008}
\bibinfo{author}{Volz, E.}, \bibinfo{author}{Heckathorn, D.D.},
  \bibinfo{year}{2008}.
\newblock \bibinfo{title}{Probability based estimation theory for respondent
  driven sampling}.
\newblock \bibinfo{journal}{Journal of official statistics}
  \bibinfo{volume}{24}, \bibinfo{pages}{79}.
\bibitem[{Vásquez-Vera et~al.(2017)Vásquez-Vera, Palència, Magna, Mena,
  Neira and Borrell}]{vasquez-vera_threat_2017}
\bibinfo{author}{Vásquez-Vera, H.}, \bibinfo{author}{Palència, L.},
  \bibinfo{author}{Magna, I.}, \bibinfo{author}{Mena, C.},
  \bibinfo{author}{Neira, J.}, \bibinfo{author}{Borrell, C.},
  \bibinfo{year}{2017}.
\newblock \bibinfo{title}{The threat of home eviction and its effects on health
  through the equity lens: {A} systematic review}.
\newblock \bibinfo{journal}{Social Science \& Medicine} \bibinfo{volume}{175},
  \bibinfo{pages}{199--208}.
\newblock \URLprefix
  \url{https://www.sciencedirect.com/science/article/pii/S0277953617300102},
  \DOIprefix\doi{10.1016/j.socscimed.2017.01.010}.
\bibitem[{{Washington State
  Legislature}()}]{washington_state_legislature_residental_nodate}
\bibinfo{author}{{Washington State Legislature}}, .
\newblock \bibinfo{title}{Residental {Landlord}-{Tenant} {Act}}.
\newblock \URLprefix
  \url{https://app.leg.wa.gov/RCW/default.aspx?cite=59.18&full=true}.

\end{thebibliography}
%


\clearpage
\appendix

\section{Data context: King County, WA}\label{data context}

King County, WA, encompassing the greater Seattle area, exemplifies the systemic issues linking eviction, housing instability, and health outcomes discussed in this paper. As the 12th-largest county in the United States, with a population of approximately 2.27 million, it is home to the third-highest number of people experiencing homelessness nationwide. Over the past decade, rapid rental market price increases, driven largely by the “Tech Boom,” have exacerbated housing insecurity and placed significant strain on social services. These trends mirror broader structural inequities, such as insufficient affordable housing, stagnant wages, and systemic discrimination, that disproportionately affect low-income households and marginalized communities.

Policy measures in King County were adopted to address these systemic drivers. Before the COVID-19 pandemic, eviction filings were relatively stable, averaging around 5,000 unlawful detainer cases annually. Recognizing the growing risk of displacement, Washington State enacted preventive policies, including extending the “pay or vacate” period from 3 to 14 days in 2019 and implementing “Just Cause” legislation in 2020, which required landlords to provide a valid reason for ending a tenancy. The state's eviction moratorium extended March 2020 through October 2021, and Seattle's specific emergency order pausing evictions ended even later, February of 2022, with more protections for renters continuing beyond state requirements. These measures marked incremental progress in addressing eviction’s structural drivers.

The onset of the COVID-19 pandemic intensified economic instability, prompting a wave of temporary protections aimed at mitigating eviction’s public health risks. The state introduced an eviction moratorium and expanded renter assistance programs, which led to a significant decline in eviction filings during the pandemic. Although many state-level protections were phased out by late 2022, local initiatives persisted. King County’s Eviction Prevention and Rent Assistance Program (EPRAP) and the Eviction Resolution Pilot Program (ERPP) provided critical support, including rental assistance and mediation between landlords and tenants. Seattle also maintained specific protections, such as a winter eviction moratorium and safeguards for households with students or school workers\footnote{\scriptsize\url{https://letherlaw.com/washington-landlords-and-tenants-remaining-covid-19-eviction-protections-lifted}}.

Despite these interventions, lifting pandemic-era protections in 2023 led to a sharp rise in eviction rates. Initially, filings remained below historical averages, reflecting the residual effects of pandemic protections. However, by July 2023, filings had returned to pre-pandemic levels, and by November, eviction rates had exceeded pre-2020 averages by over 80\%. This upward trend persisted into 2024, with sustained rates approximately 60\% above historical averages through August. These fluctuations underscore the precarious balance between temporary policy interventions and the underlying systemic pressures driving eviction.

The consequences of these trends extend beyond housing instability to broader public health concerns. King County allocates significant resources to health services for its unhoused population, including hospital care, mobile clinics, and community outreach programs. However, rising eviction rates threaten to overwhelm these systems, exacerbating health vulnerabilities among an already at-risk population. Research consistently demonstrates that preventive measures, such as rental assistance and eviction mediation, are the most effective strategies for reducing homelessness and its associated health harms \citeappendix{evans_impact_2016, garcia_many_2021}. King County’s legislative history offers valuable lessons on the potential of such measures to mitigate the adverse consequences of housing insecurity and highlights the risks of withdrawing protections prematurely.

In this context, King County serves as a critical case for understanding acute and long-term health sequelae of eviction policy. By situating our analysis within this setting, we aim to shed light on the intersection of policy, housing insecurity, and health, offering insights to inform broader efforts to address the systemic inequities underlying eviction and homelessness.

\section{Variable construction}\label{variables}

This section documents our procedures for recoding key variables used in the analysis, focusing on constructing aggregated housing status and combined race/ethnicity categories. These decisions reflect conceptual considerations and data limitations, described in detail below. Frequency distributions for all remaining variables—including those not transformed or recoded—are available in the online codebook at \url{https://uwescience.github.io/DSSG2024_understanding_homelessness/general_codebook/}.

\subsection{Housing Status}

\begin{table}[!ht]
\caption{Detailed housing characteristics of the respondents}
\label{tbl:housing}
\fontsize{12.0pt}{14.4pt}\selectfont
\begin{tabular*}{\linewidth}{@{\extracolsep{\fill}}p{2.5cm}|p{8.5cm}r@{\hskip 0.5cm}r}
\toprule
{\bfseries Housing Status} & {\bfseries Sleep Location (Last Night)} & {\bfseries N} & {\bfseries \% }\\ 
\midrule\addlinespace[2.5pt]
\multirow{2}{*}{Permanent} & Housed, other (e.g. in apartment, house) & 49 & 5.7 \\ 
 & Subsidized housing, housing vouchers & 4 & 0.5 \\ 
\midrule\addlinespace[2.5pt]
\multirow{3}{*}{Sheltered} & In an overnight shelter (e.g. mission, church, resource shelter, etc.) & 63 & 7.3 \\ 
 & In a hotel or motel & 20 & 2.3 \\ 
 & Tiny homes (not self-constructed) & 3 & 0.3 \\ 
\midrule\addlinespace[2.5pt]
\multirow{4}{*}{Temporary} & Doubled-up: couchsurfing, staying with friend or family & 113 & 13.1 \\ 
 & Maritime accommodation (e.g. boats, cargo) & 3 & 0.3 \\ 
 & Transitional, supportive, or halfway housing & 3 & 0.3 \\ 
 & Institutions such as jails, hospitals, or nursing facilities & 2 & 0.2 \\ 
\midrule\addlinespace[2.5pt]
\multirow{11}{*}{Unsheltered} & Outside in a tent (or tent-like structure) & 194 & 22.6 \\ 
 & Outside, not in a tent & 166 & 19.3 \\ 
 & In a car, truck, or van (smaller vehicle) & 81 & 9.4 \\ 
 & In an RV, trailer, or bus (larger vehicle) & 46 & 5.3 \\ 
 & In a public facility or transit (bus/train station, transit center, hospital waiting room) & 41 & 4.8 \\ 
 & In a park (uncovered, like on a bench) & 19 & 2.2 \\ 
 & Did not sleep & 18 & 2.1 \\ 
 & In an abandoned building/backyard or storage & 16 & 1.9 \\ 
 & Unsure, unknown, ambiguous & 11 & 1.3 \\ 
 & Commercial property (e.g. shops) & 1 & 0.1 \\ 
 & Workplace & 1 & 0.1 \\ 
\midrule\addlinespace[2.5pt]
\multirow{1}{*}{Missing} & Missing & 6 & 0.7 \\ 
\bottomrule
\end{tabular*}
\end{table}

\clearpage
\subsection{Race and ethnicity}

Race and ethnicity were collected using two separate survey items: one asking about racial identity and another asking whether the respondent identified as Hispanic or Latino. In reviewing the data, we observed a nontrivial pattern of missingness in these responses. Some respondents skipped multiple demographic items, consistent with typical nonresponse behavior \citeappendix{tourangeau_hard--survey_2014}. However, a cluster of respondents answered other demographic and sensitive questions (e.g., substance use, mental health) but either skipped or selected ``Do not know'' for race and/or ethnicity. We interpret this second pattern as reflecting question design effects: specifically, the absence of ``Hispanic'' as a race option may have confused respondents who primarily identify as Hispanic or Latino \citeappendix{passel_who_2024}.

Our recoding approach reflects a conceptual distinction between three types of responses: (1) affirmative identification, (2) explicit uncertainty (``Do not know''), and (3) nonresponse (``Choose not to answer'' or skip). We treat ``Do not know'' not as missing, but as a meaningful non-affirmative response, indicating that the respondent engaged with the question but did not (or could not) classify themselves using the available categories. Studies also showed that excluding ``Do not know'' responses poses a threat to validity \citeappendix{waters_dismissing_2022, denman_reducing_2018}. In contrast, skipped or refused items were coded as missing and excluded from analysis.

To construct a combined race/ethnicity variable, we followed the recent updates in the U.S. Census questionnaire \citeappendix{office_of_management_and_budget_revisions_2024} while adapting for our survey structure. Respondents who identified as Hispanic or Latino were categorized as ``Hispanic (any race),'' regardless of their race response. Non-Hispanic respondents who reported a valid race were classified accordingly. If race was answered as ``Do not know'' or left blank, but ethnicity was clearly non-Hispanic, the respondent was assigned to ``Non-Hispanic – Race not reported.'' Finally, respondents who provided neither a race nor an ethnicity response were coded as missing on the combined variable and excluded from analyses.

This strategy balances alignment with federal standards, conceptual clarity around types of nonresponse, and pragmatic concerns around retaining analytical sample size. Table~\ref{tab:race_ethnicity} presents a full breakdown of recoded categories.

\begin{sidewaystable}[p]
\fontsize{11pt}{13pt}\selectfont
\centering
\begin{tabular*}{\textwidth}{@{\extracolsep{\fill}}p{100pt}|p{200pt}p{100pt}r}
\toprule
\textbf{Race/Ethnicity Combined} & \textbf{Race} & \textbf{Ethnicity} & \textbf{N} \\
\midrule\addlinespace[2.5pt]

\multirow{3}{=}{American Indian or Alaska Native}
  & American Indian or Alaska Native & Non-Hispanic/Latino & 43 \\
  & American Indian or Alaska Native & Unknown & 1 \\
  & American Indian or Alaska Native & Missing & 2 \\

\midrule\addlinespace[2.5pt]
\multirow{1}{=}{Asian}
  & Asian & Non-Hispanic/Latino & 14 \\

\midrule\addlinespace[2.5pt]
\multirow{3}{=}{Black or African American}
  & Black or African American & Non-Hispanic/Latino & 114 \\
  & Black or African American & Do not know & 4 \\
  & Black or African American & Missing & 6 \\

\midrule\addlinespace[2.5pt]
\multirow{7}{=}{Hispanic or Latino (any race)}
  & American Indian or Alaska Native & Hispanic or Latino & 17 \\
  & Black or African American & Hispanic or Latino & 11 \\
  & Native Hawaiian or Pacific Islander & Hispanic or Latino & 6 \\
  & White & Hispanic or Latino & 41 \\
  & Do not know & Hispanic or Latino & 6 \\
  & Multiracial & Hispanic or Latino & 20 \\
  & Missing & Hispanic or Latino & 21 \\

\midrule\addlinespace[2.5pt]
\multirow{3}{=}{Multiracial}
  & Multiracial & Non-Hispanic/Latino & 86 \\
  & Multiracial & Do not know & 3 \\
  & Multiracial & Missing & 1 \\

\midrule\addlinespace[2.5pt]
\multirow{2}{=}{Native Hawaiian or Pacific Islander}
  & Native Hawaiian or Pacific Islander & Non-Hispanic/Latino & 14 \\
  & Native Hawaiian or Pacific Islander & Missing & 1 \\

\midrule\addlinespace[2.5pt]
\multirow{5}{=}{Non-Hispanic/Latino (race not reported)}
  & Do not know & Non-Hispanic/Latino & 4 \\
  & Do not know & Do not know & 6 \\
  & Do not know & Missing & 1 \\
  & Missing & Non-Hispanic/Latino & 18 \\
  & Missing & Do not know & 4 \\

\midrule\addlinespace[2.5pt]
\multirow{3}{=}{White}
  & White & Non-Hispanic/Latino & 393 \\
  & White & Do not know & 3 \\
  & White & Missing & 8 \\

\midrule\addlinespace[2.5pt]
\multirow{1}{=}{NA}
  & Missing & Missing & 12 \\

\bottomrule
\end{tabular*}
\caption{Cross-tabulation of Reported Race and Ethnicity Prior to Harmonization)}
\label{tab:race_ethnicity}
\end{sidewaystable}

\vspace{30pt}

\subsection{Eviction history}

To capture respondents’ eviction experiences, we included a short series of questions addressing both occurrence and timing. First, participants were asked: \textit{``Have you ever been evicted from a rental property?''} Those who responded ``yes'' received two follow-up questions: \textit{``Were you formally evicted by court order?''} and \textit{``How old were you when this happened?''} 

\begin{table}[ht]
\centering
\caption{Distribution of Eviction History and Type}
\label{tab:evict_type}
\begin{tabular}{llr}
\toprule
\textbf{Eviction Status} & \textbf{Formal Eviction} & \textbf{n} \\
\midrule
Non-Evicted              & --                       & 647 \\
Evicted                 & Yes                      & 98 \\
                        & No                       & 108 \\
                        & Missing                  & 1 \\
Missing & --                     & 6 \\
\midrule
\textbf{Total}           &                          & \textbf{860} \\
\bottomrule
\end{tabular}
\end{table}

Table~\ref{tab:evict_type} shows the distribution of eviction history and formal eviction status. Among 207 respondents who reported being evicted, 98 (47\%) experienced a formal, court-ordered eviction, while 108 (52\%) were evicted through informal means such as intimidation or neglect. One case was missing information on type, and six respondents were missing data on eviction history altogether.

\begin{figure}[H]
    \centering
    \includegraphics[width=0.9\linewidth]{img/evict_time_cat.png}
    \caption{Histogram of time since eviction (in years) among the evicted respondents ($n = \text{200}$). Time since eviction is calculated as the difference between current age and reported age at eviction. Seven cases were excluded due to data inconsistencies.}
    \label{fig:evict_time_cat}
\end{figure}

We calculated \textit{time since eviction} in years by subtracting the reported age at eviction from the respondent’s current age. Seven responses were excluded due to data inconsistencies in which the reported age at eviction exceeded the respondent’s current age. Figure~\ref{fig:evict_time_cat} displays the distribution of time since eviction among the analytical sample. Eviction was a relatively recent experience for most respondents: 23\% were evicted within the past 4 years, and nearly 60\% were evicted within the past 15 years. Only 14\% reported eviction more than 25 years ago. This distribution supports our interpretation of eviction as an acute or mid-term exposure for the majority of participants, rather than a distant life event.

\clearpage
\section{Methodological appendix}\label{formal}

\subsection{Formal model}

Formally, let \(A\) denote the binary treatment indicator (\(A = 1\) for individuals who experienced eviction; \(A = 0\) otherwise), \(Y\) the binary health outcome (\(Y = 1\) for adverse health; \(Y = 0\) otherwise), and \(L\) the set of covariates. The two-stage modeling framework is as follows:

\begin{enumerate}
    \item \textbf{Propensity score model:}
    \[
    \text{logit}\left( \Pr(A = 1 \mid L) \right) = \alpha_0 + \alpha_1 L
    \]
    
    \item \textbf{Weighted outcome model:}
    \[
    \text{logit}\left( \Pr(Y = 1 \mid A, L) \right) = \beta_0 + \beta_1 A + \beta_2 L + \beta_3 (A \times L)
    \]
\end{enumerate}

The inverse probability weights are defined as:

\[
w_i = \frac{A_i}{\widehat{e}(L_i)} + \frac{1-A_i}{1-\widehat{e}(L_i)}
\]

where \(\widehat{e}(L_i)\) denotes the estimated propensity score for individual \(i\).

We then used the fitted outcome model to calculate the Average Treatment Effect (ATE) via G-computation (regression standardization). Specifically, we computed predicted outcomes under two hypothetical scenarios for each individual:

\begin{enumerate}
    \item Set treatment status to treated (\(A=1\)) for all individuals and obtain predictions: 
    \[
    \widehat{\mu}_1 = \frac{1}{n} \sum_{i=1}^{n} \widehat{Y}_i^{(A=1)} = \frac{1}{n} \sum_{i=1}^{n} \text{logit}^{-1}\left(\hat{\beta}_0 + \hat{\beta}_1 + \hat{\beta}_2 L_i + \hat{\beta}_3 L_i \right)
    \]

    \item Set treatment status to untreated (\(A=0\)) for all individuals and obtain predictions:
    \[
    \widehat{\mu}_0 = \frac{1}{n} \sum_{i=1}^{n} \widehat{Y}_i^{(A=0)} = \frac{1}{n} \sum_{i=1}^{n} \text{logit}^{-1}\left(\hat{\beta}_0 + \hat{\beta}_2 L_i \right)
    \]
\end{enumerate}

The difference between these two standardized predictions provides the doubly robust plug-in estimator for the ATE:
\[
\widehat{\text{ATE}} = \widehat{\mu}_1 - \widehat{\mu}_0
\]

This doubly robust estimator remains consistent if either the propensity score model or the outcome regression model is correctly specified. 

While our implementation differs from more flexible augmented inverse probability weighting (AIPW) estimators, it shares similar double robustness properties under the canonical link function (e.g., logit) used in generalized linear models. See the supporting information of \citeappendix{gabriel_inverse_2024} for a formal discussion of this equivalence.




\subsection{Diagnostics for propensity models}

\begin{figure}[!ht]
    \centering
    \includegraphics[width=1\linewidth]{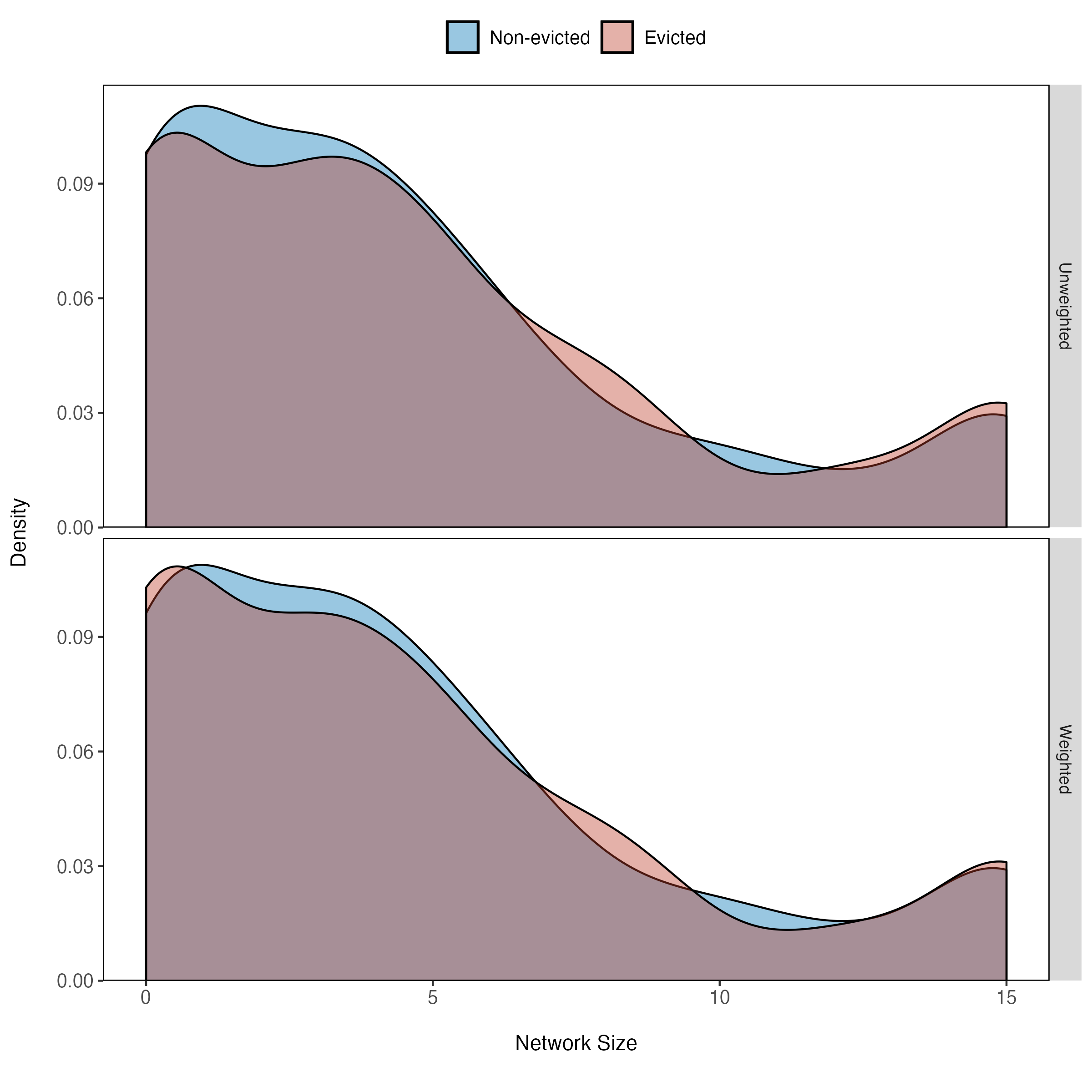}
    \caption{Distribution of network size by eviction status before and after weighting.}
    \label{fig:network_density}
\end{figure}

\begin{figure}[!ht]
    \centering
    \begin{subfigure}[t]{0.75\linewidth}
        \centering
        \includegraphics[width=\linewidth]{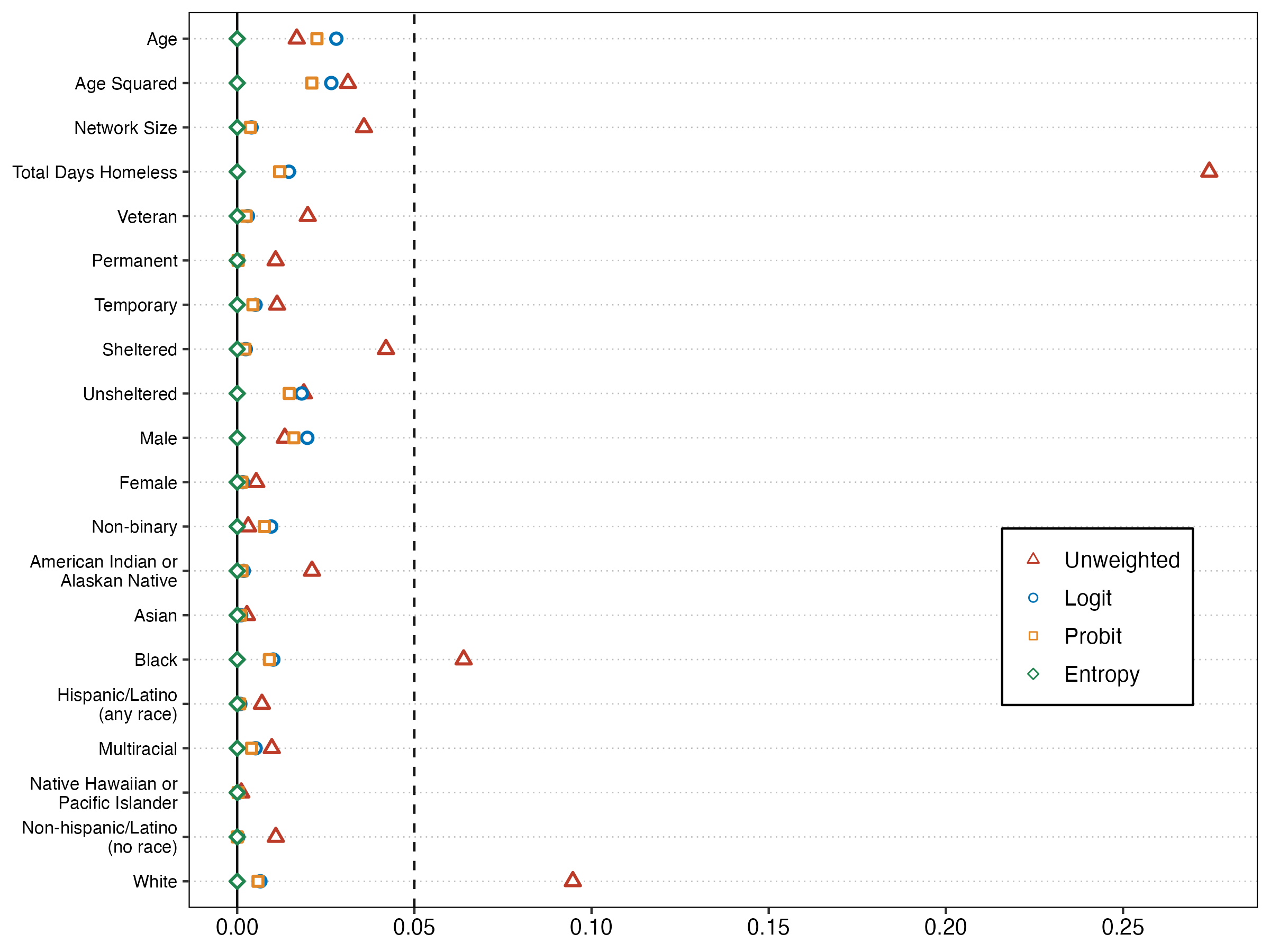}
        \caption{Absolute Standardized Mean Differences (SMD)}
        \label{fig:balance_smd}
    \end{subfigure}
    
    \vspace{0.1cm} 
    
    \begin{subfigure}[t]{0.75\linewidth}
        \centering
        \includegraphics[width=\linewidth]{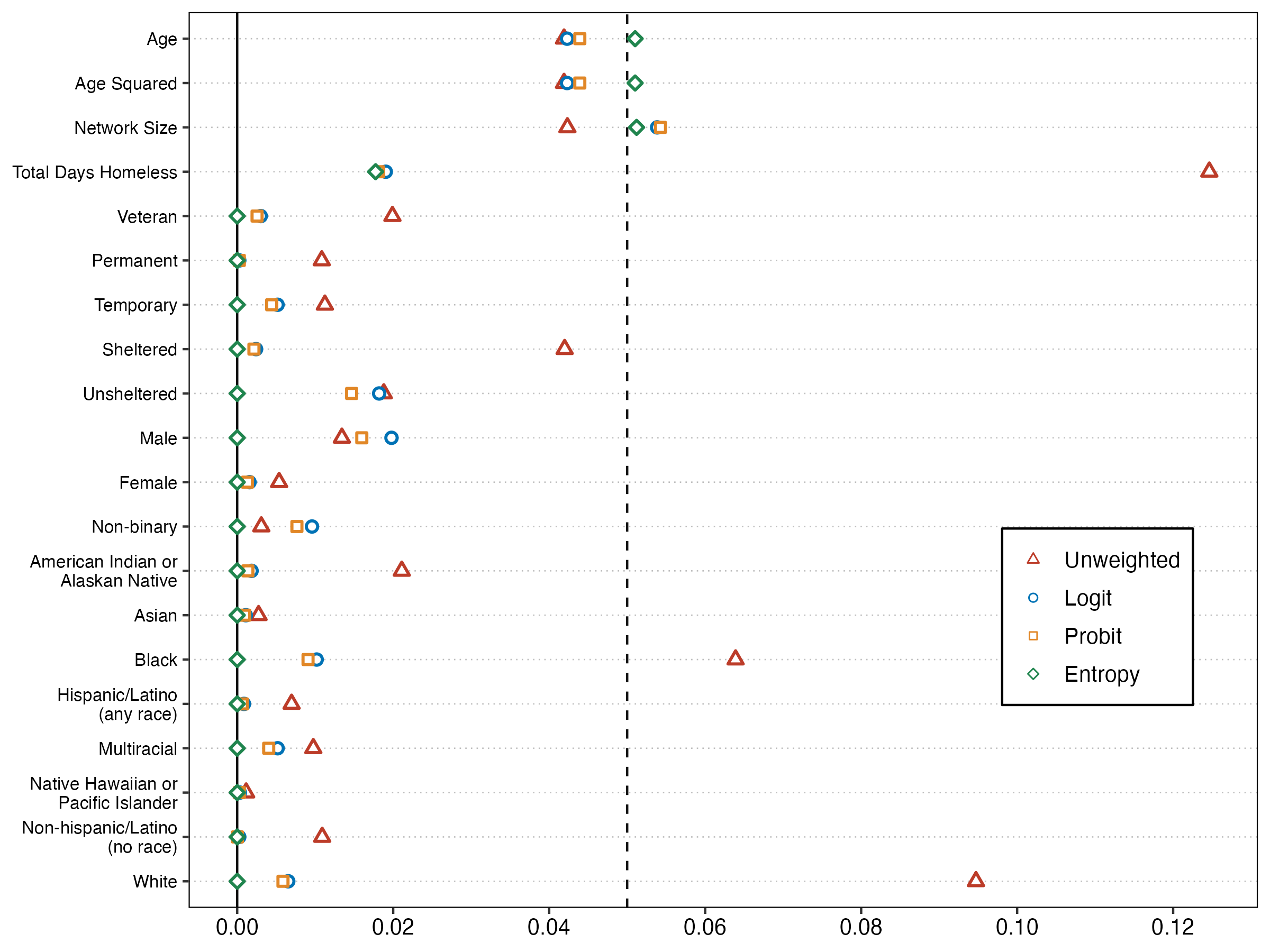}
        \caption{Kolmogorov-Smirnov (KS) Statistics}
        \label{fig:balance_ks}
    \end{subfigure}
        \caption{Covariate balance before and after propensity adjustment across two metrics.}
    \label{fig:balance_combined}
\end{figure}

Figures~\ref{fig:balance_combined}–\ref{fig:network_density} visualize key diagnostics used to assess the validity of our inverse probability of treatment weighting (IPTW) approach.

\begin{figure}[!ht]
    \centering
    \includegraphics[width=1\linewidth]{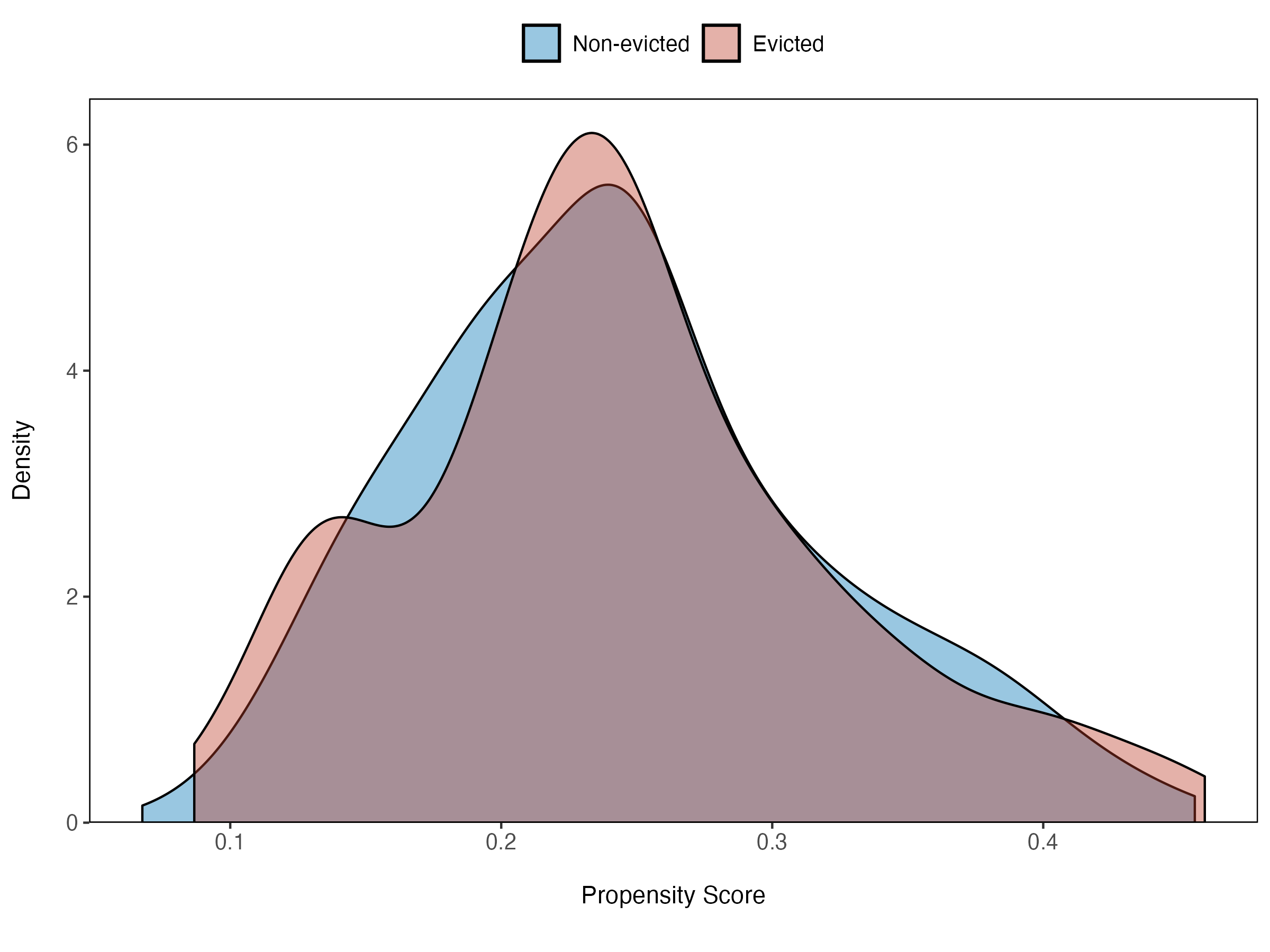}
    \caption{Distributions of estimated propensity scores by eviction status.}
    \label{fig:pscore_dist}
\end{figure}

\textbf{Figure~\ref{fig:pscore_dist}} shows the distribution of estimated propensity scores for evicted and non-evicted respondents. The substantial overlap between the two distributions indicates adequate common support, meaning that individuals in both groups had similar probabilities of eviction given observed characteristics. This is a key assumption for estimating causal effects using propensity weighting.

\textbf{Figure~\ref{fig:network_density}} presents the distribution of network size by eviction status, both before (top panel) and after (bottom panel) applying inverse probability weights. Unlike other covariates, network size distributions are already closely aligned in the unweighted data, indicating that eviction status is largely uncorrelated with network size. This is a notable finding given the respondent-driven sampling (RDS) design, where network size is typically associated with selection probability. The minimal visual change between panels confirms that the weighting procedure had limited effect on this variable, suggesting that differences in eviction status are not driven by differential visibility or network reach in the sample. This supports the assumption that treatment assignment (eviction) is independent of RDS-specific selection biases with respect to network size.

\textbf{Figure~\ref{fig:balance_combined}} displays covariate balance before and after applying propensity score weights, using standardized mean differences (top) and Kolmogorov--Smirnov statistics (bottom). In both panels, values closer to zero after weighting indicate improved balance between evicted and non-evicted groups across covariates. This suggests that the weighted sample approximates a pseudo-population where the treatment assignment is independent of observed covariates.

\clearpage

\bibliographystyleappendix{plainnat}
\bibliographyappendix{Eviction}

\end{document}